\documentclass[fleqn,usenatbib]{mnras}

\usepackage{newtxtext,newtxmath}

\usepackage[T1]{fontenc}
\usepackage{ae,aecompl}

\usepackage{xcolor}
\usepackage{graphicx}   
\usepackage{amsmath}    
\DeclareMathOperator{\sech}{sech}

\title[Pairing function of visual binaries]{Pairing function of visual binary stars}

\author[D.Chulkov]{
Dmitry Chulkov$^{1}$\thanks{E-mail: chulkov@inasan.ru}
\\
$^{1}$INASAN, Moscow, Russia\\}

\date{Accepted 2020 November 13. Received 2020 November 8; in original form 2020 September 17}
\pubyear{2020}

\begin{document}
\label{firstpage}
\pagerange{\pageref{firstpage}--\pageref{lastpage}}
\maketitle

\begin{abstract}
An all-sky sample of 1227 visual binaries based on Washington
Double Star catalogue is constructed to infer the IMF, mass ratio,
and projected distance distribution with a dedicated population
synthesis model. Parallaxes from \textit{Gaia} DR2 and
\textit{Hipparcos} are used to verify the distance distribution.
The model is validated on the single-star \textit{Tycho-2} sample
and successfully reproduces the observed magnitudes and angular
separations. The projected separation distribution follows
$f(s)\sim s^{-1.2}$ in $10^2 - 2\cdot10^3$ AU range for 1--4.5
$m_{\odot}$ primary stars. Several algorithms are explored as
pairing functions. Random pairing is confidently rejected.
Primary-constrained (PCP) and split-core pairing (SCP), the
scenarios adopting primary component's or total system's mass as
fundamental, are considered. The preferred IMF slope is $\alpha
\sim 2.8$ either way. A simple power-law mass ratio distribution
is unlikely, but the introduction of a twin excess provides a favourable result. PCP with $f(q) \sim q^{-1}$ is
preferred with a tiny twin fraction, models with $f(q) \sim
q^{-1.5}$ are acceptable when a larger twin excess is allowed. SCP
is similar to PCP when a larger slope of the power law
is adopted:  $f(q)\sim q^{\beta+0.7}$.
\end{abstract}



\section{Introduction}

Stellar mass at birth is a single parameter, which predominantly
determines future evolution and observational characteristics of a
star. The initial stellar mass distribution (IMF) is of crucial
importance, as it encodes complex star formation process,
knowledge and understanding of the IMF is essential in many areas
of astronomy. It cannot be observed directly and should be
estimated from indirect techniques \citep{2013pss5.book..115K}.
Universality of the stellar IMF and its dependence on the local
environment remains an open issue \citep{2010ARA&A..48..339B,
2014PhR...539...49K, 2014prpl.conf...53O}. Stellar multiplicity is
closely linked to the IMF in many senses \citep{2018arXiv180610605K}. It can be considered as an obstacle, creating a bias for the IMF determination, but binary systems themselves imprint valuable information on the star formation and
evolution history. Creating a theoretical model reproducing
observational properties of the multiple star population is a
challenging task yet to be solved \citep{2007prpl.conf..133G,2020SSRv..216...70L}.

The universality of stellar multiplicity is a subject of debate
\citep{2012MNRAS.427.2636K,2014MNRAS.441.3503M}. Field stars
represent a diverse population of various age formed in different
environment. There are indications that nearby star formation
regions and open star clusters are not representative of the
majority of field stars \citep{2018MNRAS.478.1825D,
2020MNRAS.496.5176D}.  Binaries with orbit size $a > 10^2$ AU are
progressively affected by the dynamical destruction according to
numerical simulations \citep{2009MNRAS.397.1577P,
2017MNRAS.465.2198D}, however field population remains indicative
of the primordial conditions \citep{2014MNRAS.442.3722P}. The
ratio of the secondary companion's mass to that of the primary,
$q=m_2/m_1$, is of particular importance, \citet{2013MNRAS.432.2378P} point out that the shape of mass ratio distribution $f(q)$ is largely unaffected by the dynamical
evolution and directly represents the outcome of star formation.
Careful interpretation and assessment of selection effects is
needed \citep{1991MNRAS.250..701T}. $f(q)$ can be considered as a
measure of component's coevolution, as an independent formation
leads to a steep decreasing $f(q)$ distribution, while mass
transfer or competitive accretion in the circumbinary disc tends
to equalize masses \citep{2017ApJS..230...15M}. This factor also
implies that $f(q)$ depends on the orbit size as the influence of
the companion is supposed to be stronger in close pairs.

The current orbit does not necessary represent conditions during
the formation of the binary system. \textcolor{black}{The
gravitational collapse of a protostellar clump followed by the
turbulent core fragmentation and subsequent rapid migration
produces companions at $\sim 10^2 - 10^4$ AU separation
\citep{2019ApJ...887..232L} and is expected to be a dominant mode
for the concerned visual binaries. Companion formation through
gravitational instabilities in the circumstellar disc produces
closer systems \citep{2016ARA&A..54..271K}. Distinct mechanisms
are responsible for the formation of wide systems. The dissolution
of star clusters appears to play a major role
\citep{2010MNRAS.404.1835K, 2011MNRAS.415.1179M} for the existence
of pairs with low binding energy. Evolution of a triple system
leading to the ejection of a companion into a distant orbit
\citep{2012Natur.492..221R} and interaction of neighbouring
protostellar cores \citep{2017MNRAS.468.3461T} are other possible
mechanisms for the formation of wide binaries.}

Binaries are ubiquitous and represent a broad spectrum of
phenomena observed in the whole electromagnetic spectrum and
beyond \citep{2017ApJ...848L..12A}. Detached unevolved systems are
normally observed as spectroscopic, visual, common proper motion,
or eclipsing binaries in optical wavelengths. Creating a large
unbiased sample representing principal properties of a stellar
population is challenging \citep{2013ARA&A..51..269D}. The
available samples are often small, leading to large error margins
and occasionally contradicting results in the literature. Below I
list several results concerning the $f(q)$ power-law slope $\beta$
for solar-mass primaries, emphasizing the large scatter of the
reported values, $q>0.2$. \citet{2009ApJS..181...62M} find $\beta
\sim -0.4$, valid for 28 -- 1590 AU separation. No variation of
$\beta=-0.5\pm0.29$ with separation is found according to
\citet{2011ApJ...738...60R}, $f(q)$ of M- and G-type stars appear
to be consistent with A- and late B-type stars.
\citet{2013A&A...553A.124R} later confirmed no dependence on
separation, but obtained $\beta=0.25\pm0.29$.
\citet{2012AJ....144..102T} suggest that the mass ratios of wide
systems ($\sim 10^4 AU$) in the solar neighborhood are distributed
nearly uniformly with a possible deficit at $q\sim 0.5$. Important
characteristic of the $f(q)$ distribution is the probable
increased frequency of equal-mass binaries ($q \sim 1$).
Originally found for close spectroscopic systems
\citep{1979AJ.....84..401L}, it is believed to be a common feature
for a broad separation range
\citep{2007A&A...463..683S,2019MNRAS.489.5822E}. The origin of
twin stars is being discussed
\citep{2000A&A...360..997T,2020MNRAS.491.5158T,2020MNRAS.494.2289A}.

In this paper, I attempt to infer the initial mass, mass ratio and
projected separation distribution evaluating an all-sky sample of
field visual binary stars with a population synthesis model.
Sections \ref{binary_sample} and \ref{model} describe the
binary-star sample and the model, its validity is tested on
single-star population in Section \ref{single_sample}. Section
\ref{results} describes the comparison of the model predictions
with observational data for binary stars, and I sum up the results
in Section \ref{conclusions}.

\section{Binary-star sample}
\label{binary_sample}
\subsection{Selection function}

The sample of binary stars is based on the \textit{Washington
Double Star Catalogue} (WDS) \citep{2001AJ....122.3466M}, which is
a principal database for visual binaries. WDS is a compilative
catalogue combining diverse observational data from various
sources. Indeed, some visual binaries are observed for several
centuries, the historic measurements remain scientifically
valuable
\citep{mayer1779novis,1782RSPT...72..112H,1893Obs....16..279L,
2013JAHH...16...81T}. Such composite dataset is obviously
non-homogeneous and it is essential to evaluate observational
biases before proceeding for further analysis.

The principal observational parameters of visual binaries are the
primary's and secondary's apparent magnitudes $mag_1$, $mag_2$
($mag_1 \leq mag_2$) and angular separation between components
$\rho$. The probability to be included in the WDS for a given
double star (the selection function) predominantly depends on
these three parameters. The WDS contains both known
gravitationally bound systems and mere projections -- the optical
pairs. Parallaxes or distances to the systems are not provided in
the WDS and, in general case, may be unknown. Below I attempt to
create a complete bias-free sample. The WDS is regularly updated,
I use the version as of August, 2019.

The WDS contains thousands of systems with subarcsecond
separation, the $\rho$ distribution reaches its peak around 0.5
arcsec. I moderately choose the lower limit as $\rho_{min}=0.8$
arcsec which reflects the resolution of the \textit{Tycho-2}
\citep{2000A&A...355L..27H} catalogue and is close to the
capability of traditional ground-based telescopes. Closer systems,
even if resolvable, may have photometry of inadequate quality,
which disturbs the further analysis. The standard form of the WDS
includes the first and the last available measurement of $\rho$
rounded to 0.1 arcsec. To reduce induced bias, I use the
exact-value version, which provides the last value of $\rho$
without rounding.

Concerning the wide binaries, physically bound systems should be
discriminated from optical pairs. Double stars with faint
companions and large separation have a larger chance to be
optical. It is possible to eliminate an optical pair if there is
information on individual parallaxes and proper motion of the
components. Another concern is a genuine physical pair with a
large $\rho$ not included in the catalogue, because it is not
deemed a double star by the observers, being too wide to be
identified among field objects. Fortunately, there are independent
datasets containing wide systems which are used to verify the
completeness of the WDS. I perform cross-identification of the WDS
with \citet{2017MNRAS.472..675A}  and \citet{2019MNRAS.489.5822E}
lists of binaries based on the \textit{Tycho-Gaia} Astrometric
Solution (TGAS) \citep{2016A&A...595A...2G} and \textit {Gaia} DR2
\citep{2018A&A...616A...1G} respectively. Particularly, I search
for the systems which are present in the referred
\textit{Gaia}-based datasets but absent from the WDS. No such
binaries are found down to the limiting magnitude 9.5 and
$\rho_{max}= 15$ arcsec. The latter value is adopted as the
maximum separation for the sample stars. Systems with $\rho \sim
1-2 $ arcsec are significantly underrepresented in
\textit{Gaia}-based datasets due to the limitations regarding
close binaries \citep{2018A&A...616A..17A}. Considering that the
maximum of $mag_1$ and $mag_2$ distributions for the whole WDS
catalogue is between 11 and 12 mag, I expect that it is complete
at least down to 9 mag for such binaries. \textit{Tycho-2}
contains around 120 thousand stars brighter than 9 mag,
corresponding to 2.9 stars per square degree or $5\cdot10^{-5}$
stars per a 15 arcsec field on average. Large contamination
from optical pairs is not expected; however later three
pairs deemed optical were excluded thanks to \textit{Gaia} DR2
data.

Another, initially unexpected, factor is related to different
magnitude systems used in the WDS. Normally, infrared or any other
non-visual magnitudes are marked by an appropriate code and can be
easily removed. However, cross-matching with \textit{Gaia} DR2
shows a significant number of stars with dramatic difference
between the WDS and \textit{Gaia} magnitudes, which is explained
with the undeclared use of red or infrared photometry in the WDS.
The number of such cases increases drastically for objects fainter
than $mag=9$, giving another reason to cutoff the sample at this
magnitude. The limit is applied both for the primary and the
secondary component. Bright stars generally are better studied
with different observational methods, there is a lower
probability for them to have undiscovered companions, so I do not
constrain the sample at the bright end.

\begin{figure}
    \includegraphics[width=\columnwidth]{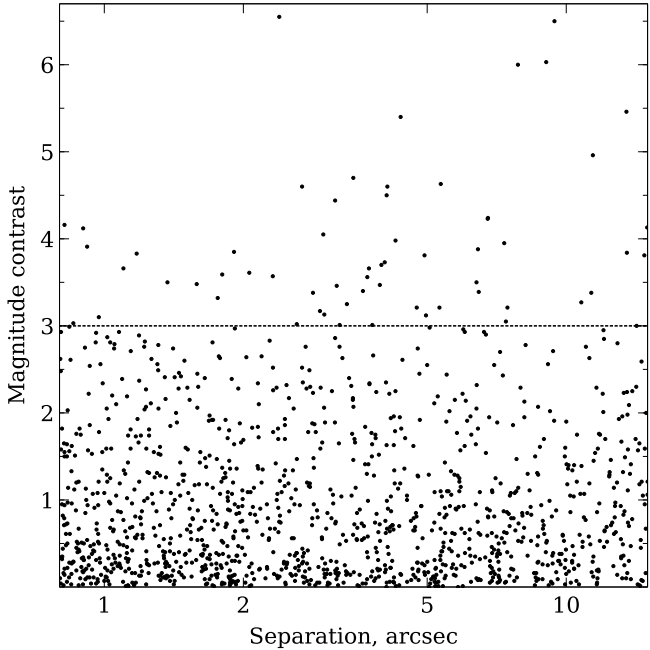}
    \caption{Observational sample binaries before cutoff of high-contrast systems, $0.8 < \rho''< 15~,~mag_{1,2}<9$. The lack of close high-contrast pairs is clearly visible. Pairs with $\Delta mag < 3$ proceed to the final sample. Sirius with $\Delta mag \sim 10$ is not shown. $\rho$ and $mag_{1,2}$ are derived from WDS, $V$ passband is expected.}
    \label{fig:sep-mag}
\end{figure}

One more factor for the selection function is the magnitude
contrast (magnitude difference) of components, $\Delta mag=
mag_2-mag_1$. Stars with similar magnitudes are more likely to be
detected in comparison to binaries with high magnitude contrast.
The extreme example of a high-contrast visual binary is Sirius,
the brightest star in the night sky, with $\Delta mag \sim 10$ and
$\rho = 10.7$ arcsec. The separation -- magnitude contrast diagram
(Fig. \ref{fig:sep-mag}) shows several systems with $\Delta mag
\sim 5-6$, but the lack of relatively close binaries with large
contrast is obvious. It seems that a close $0.8<\rho<2$ arcsec
system, even if detected, would have incorrect photometry in the
WDS. Considering Fig. \ref{fig:sep-mag}, $\Delta mag < 3$ is set
as an appropriate limit applied for all separations. Summing up,
the following conditions set up the selection function:

\begin{equation}
    \label{eq:selection}
0.8^{\prime\prime} < \rho < 15^{\prime\prime} ~,~ mag_{1,2}<9 ~,~ \Delta mag < 3
\end{equation}

\subsection{Multiple systems \textcolor{black}{\& excluded entries}}
A significant fraction of binaries are parts of multiple systems.
For clarity, here I call a system multiple if it consists of three
or more components. Various volume and magnitude-limited samples
\citep{2008MNRAS.389..869E,2010ApJS..190....1R,2014AJ....147...87T}
show $\frac{N_{\rm Mult}}{N_{\rm Bin}+N_{\rm Mult}} \sim 20-30$
per cent fraction of multiples. In general, it may be possible to
exclude known multiple systems and keep only pure binaries in the
sample. I opt not to do so for two reasons. First, such approach
is certainly biased, as higher-order multiplicity is
characteristic for massive stars; leaving them out significantly
distorts the sample. The second reason is that systems considered
as binaries may have yet undiscovered companions. Therefore, I
attempt to reduce multiple systems to binaries rather than exclude
them. There is certainly more than one way to do it, making
treatment of multiple stars one of the sources of ambiguity in the
present study.

The following receipt is adopted. When the WDS contains several
entries with the same designation, Eq. \ref{eq:selection}
conditions are checked. If only one pair is suitable, it proceeds
to the sample alone. The most common case is that the component is a close binary on its own, for example:

WDS 00174+0853 AB, $\rho= 0.162^{\prime\prime}$ , 8.38+7.78 mag

WDS 00174+0853 AB, C, $\rho= 3.953^{\prime\prime}$, 7.13+7.66 mag
\\Here I ignore the AB pair, since its separation is less than the 0.8 arcsec limit and keep AB,C pair in the sample.

WDS 18029+5626 AB, $\rho= 35.909^{\prime\prime}$, 7.78+8.14 mag

WDS 18029+5626 AC, $\rho= 33.592^{\prime\prime}$, 7.78+8.53 mag

WDS 18029+5626 BC, $\rho= 6.036^{\prime\prime}$, 8.14+8.53 mag

In the example shown above, all the three possible combinations for a triple system are present in the WDS. Pairs AB and AC are too wide ($\rho> 15$ arcsec), so pair BC is included into the sample, while AB and AC are ignored.

Occasionally, several entries meet the criteria for component
brightness and separation. Duplicate inclusion of a star in the
sample is undesirable; in such cases, the pair with brighter
companions is selected. In the next example, both entries meet the
$\rho$ and $mag_{1,2}$ conditions, AC is chosen since its
secondary companion is brighter than in the AB pair:

WDS 10441-5935 AB, $\rho= 2.000^{\prime\prime}$, 8.59+8.64 mag

WDS 10441-5935 AC, $\rho=13.670^{\prime\prime}$, 7.89+8.59 mag

For several higher-order multiples (WDS 01158-6853, 05353-0523 ,
05381-0011, 16120-1928, 18443+3940) it is possible to keep both AB
and CD pairs, each component is included into the sample only
once.

Stellar multiplicity is not limited to resolved binaries, which
are contained in the WDS. Unresolved multiplicity can be
suspected from spectroscopic observations or stellar variability.
I do not aim for a rigorous survey of stellar multiplicity and
admit that some stars in the sample are close detached,
semidetached, or contact binaries. If a component of the visual
binary is an unresolved or close visual binary, the WDS usually
provides integral brightness of the close pair. In the particular
case of equal companions, the integral brightness is 0.75 mag
higher than that of an individual star. If the secondary companion
is relatively faint, the difference is less significant. There are
few cases when the WDS provides the resolved primary's magnitude
for a close binary instead of the integral brightness or when
magnitudes for the components of a multiple system are
contradictory. I choose not to alter the WDS magnitudes and treat
them as is.

The WDS is primarily an astrometric catalogue, photometry is not
in its prime focus, notably it lacks sources of the provided
magnitudes. Cross-matching shows that, considering the sample, the
absolute majority of stellar magnitudes come to the WDS from
\textit{Tycho-2}. \textit{Tycho-2} \citep{2000A&A...355L..27H} is
a dedicated homogeneous photometric catalogue, but it does not
cover the sample completely, as many stars, predominantly close
and faint companions, are missing. \textit{Catalog of Components
of Double} \& \textit{Multiple stars} (CCDM)
\citep{2002yCat.1274....0D} is another important source of
magnitudes in the WDS. Photometry of visual binaries is
complicated and discrepancies in different datasets are
anticipated. This topic is not widely covered in the literature, I
refer to \citet{2005A&A...431..587P}, who have found a bias around
0.1 mag between speckle interferometry estimates and
\textit{Hipparcos} \citep{1997A&A...323L..49P} values. Due to the
absence of a better homogeneous dataset, I stick to the WDS data.

The considered sample with $mag_{1,2} < 9$ is at the bright end of
\textit{Gaia} DR2 \citep{2018A&A...616A..17A}. \textit{Gaia} data
are incomplete in this magnitude range and not suited to
completely replace the WDS-based sample, still it is useful for
the verification of certain systems and provides parallaxes for
most of them. The WDS includes some spurious systems, considering
\textit{Gaia} DR2 data, CCDM catalogue, and sky imagery in
\textit{Aladin} \citep{2014ASPC..485..277B}, I have omitted all
systems marked `X' (`Dubious Double') and `K' (infrared
magnitude). Several systems with the last observation before 1991,
zero number of measurements, or $mag_2$ unavailable in the WDS
were also excluded. Additionally, the following systems were
excluded due to their incorrect magnitudes:

WDS 10396-5728 is a wide pair with equally bright companions
according to the WDS ($\rho= 8.6$ arcsec, 8.46+8.4 mag); however
2MASS \citep{2006AJ....131.1163S} shows a much fainter secondary
star, consistent with \textit{Gaia} $G=12.2$ mag.

WDS 15365+1607 ($\rho= 1$ arcsec, 8.5+8.9 mag) is not identified.
Magnitudes are rounded, which is often an attribute of infrared
photometry in the WDS. The coordinates refer to the high proper
motion star 18~Ser of 5.9 mag, no reference on its multiplicity is
found.

WDS 17404-3707 is another wide ($\rho= 10.3$ arcsec) system with
components of 8.3 mag according to the WDS; however, the 2MASS
image clearly shows a fainter secondary companion in agreement
with \textit{Gaia} $G \sim 13$ mag.

The WDS 21308+4827 identifier matches two systems with almost
identical $\rho$ and positional angles, but different $mag_2$.
\textit{Gaia} DR2 confirms that $mag_2$ is around 11 mag.

WDS 23293-8543 has $mag_2=10.9$ both in the CCDM and \textit{Gaia}
DR2, which contradicts $mag_2=8.53$ in the WDS.

Several other systems show magnitudes in the CCDM or \textit{Gaia}
DR2 slightly fainter than $mag=9$. I do not alter them on a
case-by-case basis and keep the original WDS data. Additionally
three pairs are removed from the sample as optical. These are WDS
17150+2450, 12095-1151, and 17150+2450, wide pairs
with linear solutions according to the WDS. The \textit{Gaia} DR2
parallax and proper motion data confirm that these double stars
are not bound.

\subsection{Distances and general sample properties}
\label{parallaxes}

The final sample contains 1227 binaries. \textit{Gaia} DR2 has a
resolved $G$-magnitude photometry for both companions in 1107 cases. Distance estimates from \citet{2018AJ....156...58B} for
both components are available for 1032 systems; for 155 binaries
it is derived for just one companion. This leaves 40 entries
without \textit{Gaia} DR2 distances, inverse parallaxes $d=1/\pi$
from \textit{Hipparcos} \citep{2007A&A...474..653V} are used for
them. The only system without \textit{Gaia} or \textit{Hipparcos}
parallax is WDS 08068-3110 (HD 67438). The B9V primary star is
expected to have 0.7 mag absolute $V$ magnitude
\citep{2013ApJS..208....9P}, which implies $d=363$ pc, absorption
neglected. The system is located near the Galactic equator,
allowing interstellar extinction around 0.5 mag, I estimate its
distance around 290 pc. Thereby, beside $\rho$ and $mag_{1,2}$,
distance estimates are available for all binaries in the sample.
Since individual parallaxes are prone to significant errors,
quartile and median values are used for comparison with the model
predictions. Distance estimates for both companions provide a
natural estimate of observational uncertainties.

It is important to acknowledge that \textit{Gaia} algorithms treat
all stars as single, whereas the sample comprises stars certainly
known to be binary or multiple. However, the majority of the
sample binaries have orbital periods large enough to be negligible for \textit{Gaia}. A comprehensive analysis of
\textit{Gaia} data is beyond my scope, but it is evident that
\textit{Gaia} DR2 contains inaccurate or even wrong
data for some of the objects. Indeed, for 138 sample binaries, the
primary and secondary components' parallaxes fall beyond their
three-sigma error. There is a possibility that some of them are
optical pairs, but in most cases, this discrepancy
reflects \textit{Gaia} systematic errors rather than actual
difference in the distance. This uncertainty discourages me from
creating a volume-limited sample. The furthest binary with
well-agreeing components' parallaxes is WDS 10441-5935 at $d \sim
3500$ pc. For at least 5 binaries, estimates for one of the
components point out to the 4--6 kpc distance range.

\begin{figure}
    \includegraphics[width=\columnwidth]{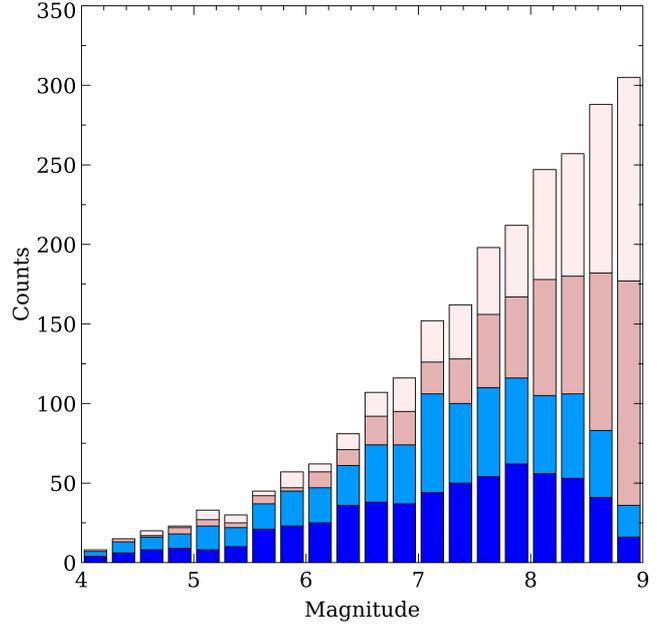}
    \caption{Apparent magnitude \textcolor{black}{($V$)} distribution for the 1227 sample binaries. The areas refer to (bottom up) primaries of close ($0.8 < \rho''< 2.785 $) and wide ($2.785<\rho''<15$) \textcolor{black}{pairs}, secondaries of close and wide \textcolor{black}{pairs}. Close and wide populations look similar.}
    \label{fig:mag_wide_close}
\end{figure}

\begin{figure}
    \includegraphics[width=\columnwidth]{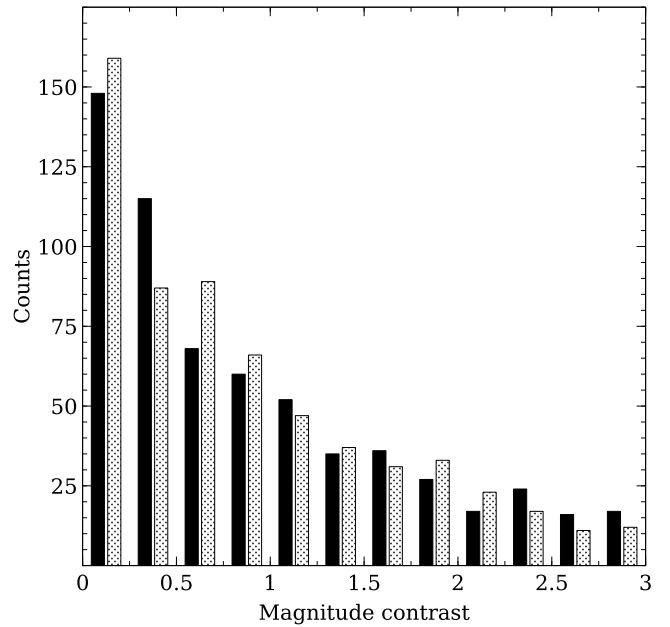}
    \caption{\textcolor{black}{$V$ band} magnitude contrast $mag_2-mag_1$ distribution for close (dark color) and wide sample binaries. Close pairs prevail among high-contrast systems contrary to the potential bias.}
    \label{fig:delta_wide_close}
\end{figure}

The median $\rho$ of the sample binaries is 2.78 arcsec. Fig.
\ref{fig:mag_wide_close} and \ref{fig:delta_wide_close} show
distributions of $mag_{1,2}$ and $\Delta mag$ for the close and
wide binaries separately. No evident difference potentially caused
by the observational biases or sample incompleteness is
noticeable. Statistical tests show a high p-value suggesting that
$mag_{1,2}$ and $\Delta mag$ are drawn from the identical
distribution for close and wide binaries.

\section{Population synthesis model}
\label{model}
\subsection{Overview and general assumptions}
The stellar population model is based on a series of Monte Carlo experiments exercising random sampling of astrophysical parameters from prior probability distributions. Observable parameters
$mag_{1}$, $mag_2$ and $\rho$ are calculated from  astrophysical characteristics: stellar masses, ages, distances, etc. Then the selection function (Eq \ref{eq:selection}) is
applied and predictions are compared to the observational sample in Section \ref{single_sample} and \ref{results}. I briefly
explore how varying specific parameters affects the outcome in Section \ref{modified_discuss}. This section describes general
assumptions, relations, and adopted distributions of the standard model.

Although studying binaries is the main goal of the present paper,
the model is used to reproduce single-star population as well,
mainly for the purpose of validation. This model does not consider
higher-order systems and multiplicity frequency, it creates either
entirely single or binary synthetic population. Another important
simplifying assumption is that stars evolve as isolated objects.
Sample binaries indeed have wide orbits, the mutual influence of
the components is expected to be negligible. However, if the
companion is a close binary on its own, its evolution may be
disturbed. Luminosity or absolute magnitude of an isolated star
depends on its mass, age, and metallicity:
$L=Mag_{1,2}=f(m,t,[Fe/H])$. The total number of generated objects
in the model is chosen arbitrary and essentially is a compromise
between computational time and convergence of the simulation
results. The size of the synthetic binary sample is chosen to be
at least 20 times larger than the observational dataset comprising
1227 objects.

The following underlying parameters are generated to predict
observable parameters of the sample:
\begin{itemize}
\item Stellar mass at birth $m_1$ and $m_2$
\item Stellar age $t$
\item Metallicity $[Fe/H]$
\item Galactic coordinates and distance $d$
\item Interstellar extinction $A_v$
\item Projected (linear) distance between components $s$
\end{itemize}

\subsection{IMF and pairing functions}
\label{IMF_f(q)}

The distribution of primordial stellar masses is determined by the
initial mass function (IMF). Power law is commonly used since
pioneer research of \citet{1955ApJ...121..161S}. Popular forms of
IMF used nowadays are a broken power law
\citep{2001MNRAS.322..231K} and a combination of log-normal
distribution with a power law \citep{2003PASP..115..763C}.
Different analytical forms exist; however, their outcomes are
essentially indistinguishable \citep{2013MNRAS.429.1725M}.
Therefore, I do not attempt to discriminate them and adopt a
multiple-part power law. The initial stellar mass is drawn from
the probability density function:
\begin{equation}
    \label{eq:imf}
\dfrac{dN}{dm} \sim \begin{cases} 0~,~ m<m_{\rm min}~, ~m>m_{\rm max} \\
m^{-\alpha_1}~ ,~ m_{\rm min} \le m<m_{\rm split}\\
m^{-\alpha_2}~ ,~ m_{\rm split} \le m \le m_{\rm max}.\\
\end{cases}
\end{equation}

Low-mass stars have low luminosities, and their contribution to
the sample containing relatively bright stars is very limited.
Empirically, $m_{\rm min}=0.3 m_{\odot}$ is chosen as a cutoff,
lower-mass stars virtually never get to the synthetic sample, but
still take computational time. The selection of $m_{\rm split}$ is
flexible, by default $m_{\rm split}=0.5 m_{\odot}$. $m_{\rm
max}=68 m_{\odot}$ is determined by the availability of
isochrones, but the contribution of the highest-mass stars is
efficiently limited by their short lifespan anyway. Addition of a
larger number of power-law segments is possible, but practically
not necessary. $\alpha_1=1.3$ is used by default, while a wider
range of $\alpha_2$ is explored.

IMF in Eq. \ref{eq:imf} allows to draw a single-star
population, more assumptions are needed for binaries. Various ways
of pairing stars into binaries are extensively covered in
\citet{2009A&A...493..979K}. Different scenarios represent
characteristics that are deemed to be fundamental in a particular
algorithm. The four principal parameters for binaries are the
primary's and secondary's masses $m_1$ and $m_2$,  $m_1 \ge m_2$,
the total mass $m_T=m_1+m_2$, and the mass ratio
$q=\frac{m_2}{m_1}~, ~0<q\le 1$ \footnote{These definitions apply
to the model description only. If the lower-mass companion becomes
the brighter one in a pair, it is considered as the primary during
comparison of the model predictions against observational data.}.
Several pairing mechanisms are considered. The most
straightforward one is to assume that companions' masses are both
independently drawn from the same IMF. The star that happens to be
more massive in the pair is declared primary, the other one
becomes secondary, $m_1\ge m_2$. Such algorithm or pairing
function is known as random pairing (RP). While RP produces very
steeply decreasing $f(q) \sim q^{-\alpha_2}$ distribution when a
restricted subset of massive stars is considered
\citep{1961PASP...73..439W}, the overall distribution is actually
growing as the sample is dominated by low-mass stars
\citep{1991A&A...247...87P}.

\begin{figure}
    \includegraphics[width=\columnwidth]{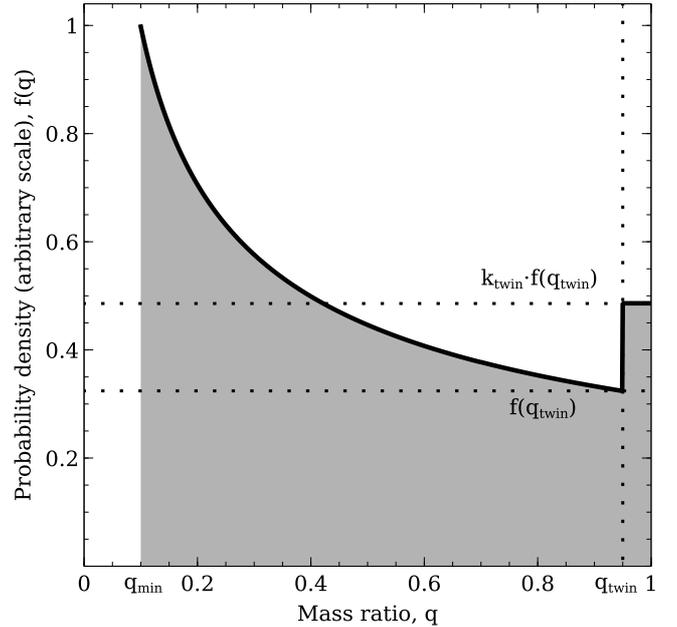}
    \caption{Mass ratio ($q=m_2/m_1$) probability density function $f(q)$ with twin excess. $q \sim q^\beta$ for $q_{\rm min} \le q<q_{\rm twin}$, $q_{\rm twin}$ is always set to 0.95; in the shown example $\beta=-0.5$, $k_{\rm twin}=1.5$.}
    \label{fig:q_twin}
\end{figure}

Other pairing functions imply that the components' masses are
correlated and explicitly depend on the mass ratio. Similarly to
the IMF, the broken power law $f(q)=\dfrac{dN}{dq} \sim q^{\beta}$
is normally used to define the distribution, $\beta$ may vary as a
function of $q$. Systems with low $q$ are unlikely to pass through
the selection function (Eq. \ref{eq:selection}), making the sample
not sensitive to such binaries; the use of a different $\beta$ in
the low-$q$ segment is unnecessary, as the universal $\beta$ covers
the whole $q$ range. There is evidence suggesting an excess of
binaries with $q \sim 1$, therefore an additional twin excess is
introduced to accommodate binaries with the companions of similar
mass. The distribution of q in the $q_{\rm twin}\le q \le 1$ range
is considered uniform, the probability density is multiplied by
the factor $k_{\rm twin}$, see Fig. \ref{fig:q_twin}.
\textcolor{black}{The twin excess factor is derived as the ratio of
one-sided $f(q)$ limits at $q=q_{\rm twin}$}: $k_{\rm twin}=
\dfrac{\lim_{q\to q_{\rm twin}^+}f(q)}{\lim_{q\to q_{\rm
twin}^-}f(q)}$ \textcolor{black}{and expresses the relative surplus
of generated pairs with nearly identical companions}.

\begin{equation}
    \label{eq:q}
f(q) = \dfrac{dN}{dq} \sim \begin{cases} 0, ~q<q_{\rm min}~, ~q>1 \\
q^{\beta}, ~q_{\rm min}\le q<q_{\rm twin}\\
1, ~q_{\rm twin}\le q\le1, \frac{f(q_{\rm twin}+)}{f(q_{\rm twin}-)}=k_{\rm twin}\\
\end{cases}
\end{equation}

$q_{\rm min}$ is set to 0.1; as already mentioned, low-$q$ systems
almost never pass through the selection function; $q_{\rm
twin}=0.95$, as in \citet{2017ApJS..230...15M}.  They
define the twin excess as the ratio $f_{\rm twin}$ of additional
twin stars with $q>0.95$ relative to the total number of binaries
with $q>0.3$; this value is calculated along with $k_{\rm twin}$.
$f(q)$ without the twin excess is also considered, then
$f(q) \sim q^{\beta}$ for $~q_{\rm min} \le q \le 1$.

\begin{figure}
    \includegraphics[width=\columnwidth]{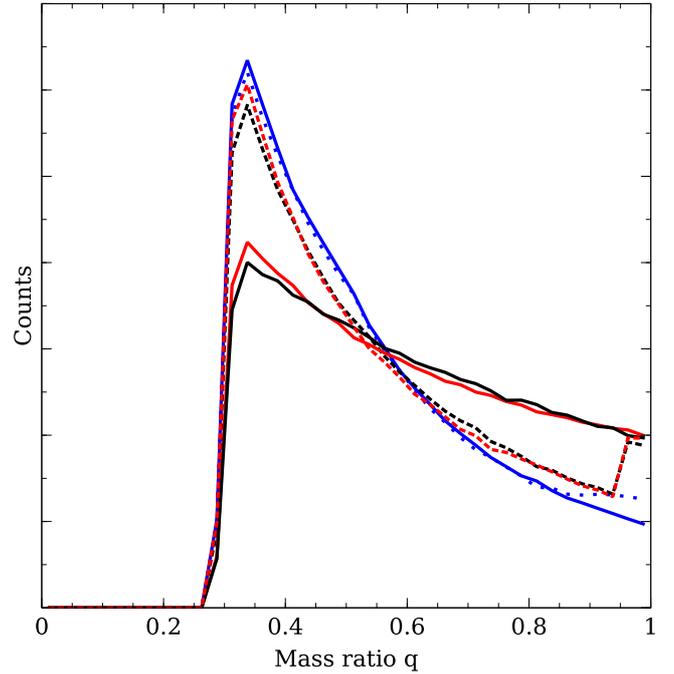}
    \caption{Produced $f(q)$ distributions for the subsample of solar-mass primaries $0.95m_{\odot}<m_1<1.05m_{\odot}$. $a_2=2.8$ is adopted for PCP and SCP. Red solid-- PCP, $\beta=-0.7$; black solid-- SCP, $\beta=0$; red dotted-- PCP, $\beta=-1.5$, $k_{\rm twin}=1.5$; black dotted-- SCP, $\beta=-0.8$, $k_{\rm twin}=1.5$; blue solid -- RP, $\alpha_2=2$, blue dotted -- RPT, $\alpha_2=2$, $f_{\rm twin}=0.04$. The distributions for PCP and SCP are nearly identical when $\beta$ larger by 0.7 is used for SCP. RP and RPT produce a steep $f(q)$ distribution. }
    \label{fig:PCP_SCP_RP}
\end{figure}

The pairing functions requiring $f(q)$ are primary-constrained pairing (PCP) and split-core pairing (SCP). PCP assumes that the primary mass $m_1$ is drawn from IMF; then, the $f(q)$ distribution is used to assign the mass ratio. The secondary mass is calculated as $m_2=q\cdot m_1$. Another option is
SCP, which adopts IMF to draw the system's total mass
$m_{T}=m_1+m_2$ and $f(q)$ for the mass ratio. Individual masses
are calculated as $m_1=\frac{m_T}{1+q} ~,~ m_2=m_T-m_1$. Both for
PCP and SCP, systems with $m_2<m_{\rm min}$ are removed. IMF (Eq.
\ref{eq:imf}) is adapted for SCP, $m_{\rm min}$, $m_{\rm split}$,
and $m_{\rm max}$ are multiplied by a factor of 2, as the total
mass of the system is drawn, rather than that of an individual
star. \citet{2013MNRAS.430L...6G} noticed that SCP with flat $f(q)
\sim q^0$ distribution was roughly similar to PCP with $f(q) \sim
q^{-0.5}$. My simulations confirm the relation; however, the size
of correction depends on the IMF slope, SCP with $f(q) \sim
q^{\beta}$ produces a distribution almost identical to PCP with
$f(q) \sim q^{\beta-0.7}$, if $\alpha_2=2.8$, see Fig.
\ref{fig:PCP_SCP_RP}. The relation remains valid for models with the twin excess and obstructs the choice between PCP and SCP.

\begin{figure}
    \includegraphics[width=\columnwidth]{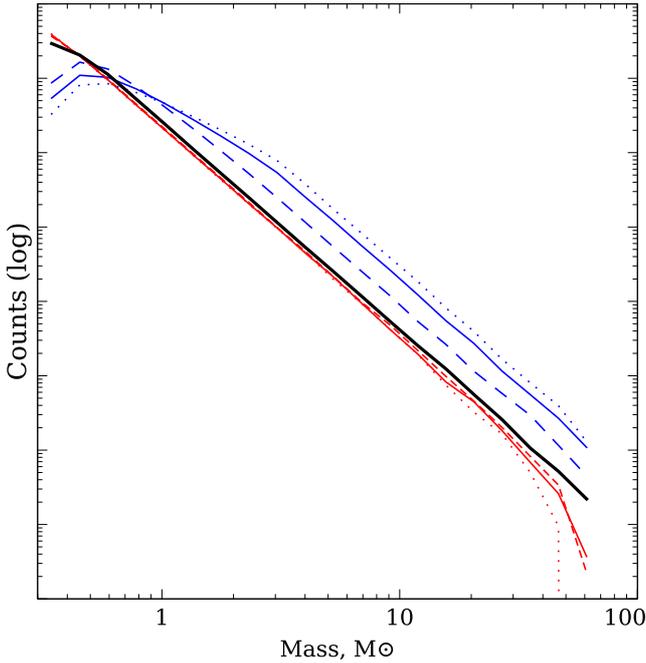}
    \caption{Generating and produced mass distribution for PCP pairing mechanism with different $f(q)$. The black solid curve is the generating IMF, $\alpha_2=2.8$ in Eq. \ref{eq:imf}. The dashed curve: $f(q) \sim q^0$; dotted curve: $f(q) \sim q^{-2}$; colored solid curve: $f(q) \sim q^{-1.5}$ with twin excess $k_{\rm twin}=1.5$. Distribution of primary components $m_1$ is shown in blue; red is reserved for secondary components $m_2$. Although the shape of the produced distribution depends heavily on $m_{\rm min}$ and $q_{\rm min}$, properties of the produced synthetic sample remain unaltered until these parameters are small enough.}
    \label{fig:PCP_mass}
\end{figure}

\begin{figure}
    \includegraphics[width=\columnwidth]{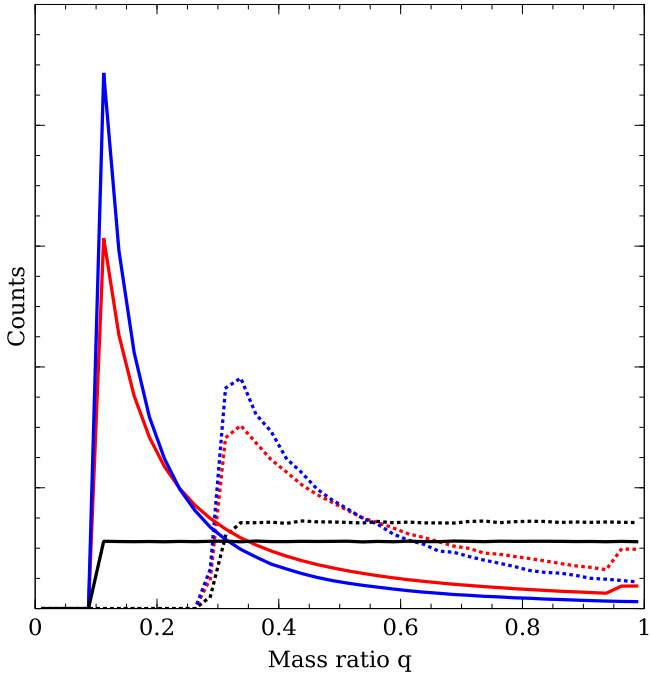}
    \caption{Generating and produced mass ratio distribution for PCP with different $f(q)$. Solid curve: generating distribution; dotted curve: produced distribution for a subsample of solar-mass primaries $0.95m_{\odot}<m_1<1.05m_{\odot}$. Color refers to various $f(q)$: black $\sim q^0$, blue $\sim q^{-2}$, red $\sim q^{-1.5}$ with twin excess $k_{\rm twin}=1.5$.}
    \label{fig:PCP_q}
\end{figure}

Both PCP and SCP imply that $f(q)$ does not depend on stellar mass. For instance, the same f(q) is used to generate $m_2$ for the primary masses $m_1=m_{\odot}$ and $m_1=5m_{\odot}$. Of course, scenarios with $f(q)$ varying as a function of mass are possible and potentially may better match astrophysical processes, rather than the simple approach chosen here. The equations
\ref{eq:imf} and \ref{eq:q} should be considered as the generating mass and mass ratio functions. The produced distribution is
different, as the systems with low-mass companions $m_2<m_{\rm min}$ are removed from the sample. For instance, $m_1= 0.5m_{\odot}$ primary star happens to be generated in PCP scenario. If the drawn
value of mass ratio is less than $q=\dfrac{m_{\rm
min}}{m_1}=\dfrac{0.3m_{\odot}}{0.5m_{\odot}}=0.6$, the system would be eliminated. Hence, effectively, $f(q)=0$ for $q<0.6$ for
a half-solar-mass primary. The elimination of systems with low-mass companions transforms the generated distributions, see Fig. \ref{fig:PCP_mass} and \ref{fig:PCP_q}. The shape of the produced functions depends heavily on $m_{\rm min}$ and $q_{\rm
min}$, none the less simulations produce  identical synthetic populations
until $m_{\rm min}$ and $q_{\rm min}$
are small enough, reaffirming that the generating distributions
determine the outcome of the model. Note that the overall $f(q)$
distribution differs significantly from the $f(q)$ produced for a specific mass. Again, I refer to \citet{2009A&A...493..979K} for a
complete review of pairing scenarios.

In addition to RP, PCP, and SCP, I consider distinct random
pairing + twins scenario (RPT). Under this algorithm, masses $m$ are
initially assigned according to the IMF, Eq. \ref{eq:imf}. Then, most stars are paired through normal random pairing. However, the small fraction of stars $f_{\rm twin}$ forms a distinct twin
population. Masses of these twin stars are divided according to
normal distribution  $f(Q)\sim N(\mu=0.5, \sigma=0.02)$, $m_1=Q
m$, $m_2=m-m_1$. $m_1$ and $m_2$ may be switched to ensure $m_1
\ge m_2$. The resulting median $q=m_2/m_1$ for the population of
twin stars is around 0.95, matching $q_{\rm twin}$ used for the
PCP and SCP algorithms.  Normal random pairing can be considered
as RPT with $f_{\rm twin}=0$. The produced $f(q)$ distribution for
RP and RPT scenarios is shown in Fig. \ref{fig:PCP_SCP_RP}.

\subsection{Age and metallicity distribution}
\label{age_and_metal}

The model assumes that the stellar IMF does not change over time, stellar ages are drawn independently from masses. A constant star
formation rate (SFR) is often used as a first approach; however, it is not able to reproduce accurately the \textit{Tycho-2} data
and decreasing SFR is favoured \citep{2014A&A...564A.102C}. Here I
adopt a simple exponential law:
\begin{equation}
    \label{eq:sfr}
\dfrac{dN}{dt} \sim \exp{\gamma t}
\end{equation}

Considering table 7 from \citet{2009MNRAS.397.1286A}, I set
$\gamma=0.1~Gyr^{-1}$, $t_{\rm max}=13$ Gyr, current SFR is at 27
per cent of the initial level. The minimum age is determined by
availability of stellar isochrones, $t_{\rm min}=4\cdot10^6$
years.

The last parameter required to set the stellar luminosity is metallicity. It depends on the stellar age, as older population tends to have lower $[Fe/H]$ with a larger scatter \citep{2000A&A...358..850R}. $[Fe/H]$ is adopted normally distributed with the mean and the standard deviation being a function of stellar age, see Table \ref{tab:amd}. There are indications that binary-star properties are correlated with metallicity \citep{2019ApJ...875...61M}, however, low-metallicity stars are scarce in the sample and their influence is limited. Age and metallicity are assumed identical for both companions since the formation of binary by a capture is rare \citep{1991MNRAS.249..584C}.

\begin{table}
    \centering
    \caption{Parameters of normal distribution for metallicity adopted in the model, derived from fig. 13c of \citet{2006MNRAS.371.1760H}. For details, see Section \ref{age_and_metal}.}
    \label{tab:amd}
    \begin{tabular}{ccc}
        $\overline{Fe/H}$ & $\sigma$ & ages \\
        \hline
        0 & 0.1 & t < 2 Gyr\\
        -0.1 & 0.1 & 2 < t < 4 Gyr\\
        -0.1 & 0.15 & 4 < t < 8.5 Gyr\\
        -0.2 & 0.2 & 8.5 < t < 10 Gyr \\
        -0.5 & 0.3 & t > 10 Gyr \\
        \hline
    \end{tabular}
\end{table}

After stellar mass, age, and metallicity are drawn, they are
converted to luminosity with a set of isochrones produced by
PARSEC 1.2S \citep{2013ApJS..208....9P, 2014MNRAS.444.2525C,
2014MNRAS.445.4287T, 2015MNRAS.452.1068C, 2017ApJ...835...77M,
2019MNRAS.485.5666P}. The \textit{Tycho-2} $V_T$ band is used since most sample stars have photometry in this system; for single stars, $B_T$
magnitude is additionally calculated. The grid of isochrones
covers stellar ages in the $6.6 <\rm{\log}~ t~(Gyr)< 10.11$ range
and metallicities in the $-1.6<[Fe/H]<0.6$ interval, with the step
0.01 and 0.1 dex, respectively. The nearest available value is
selected for the stellar age and metallicity, linear interpolation
is used for mass. All evolutionary sections in PARSEC are included
with the exception of the final post-AGB stage; degenerated stars
are absent both from the sample and the model. The IMF transforms
to the present-day mass function at this stage, as the stars older
than their evolutionary limit are eliminated.

\subsection{Spatial distribution and extinction}
\label{distance} Next, it is necessary to constrain the spatial
distribution. The Milky Way has a complex structure
\citep{2016ARA&A..54..529B}, any model is an inevitably simplified
representation of the real Galactic stellar population. With the
focus on bright and relatively close stars, a model consisting of
the thin and thick disc is chosen. The Galactic halo or bulge are
not incorporated since their input is negligible.
The distinction between the thin and thick disc is defined by the
stellar age, stars with $t > 10$ Gyr are considered to be a part
of the thick disc, while younger stars are treated as a part of
the thin disc. Simple exponential laws $\dfrac{dN}{dr} \sim
\exp{\dfrac{-r}{L}}$  are used for the radial distribution with
the scalelength $L=2500$ and $L=3500$ pc for the thin and thick
disc, respectively, \textcolor{black}{$r$ is a projected distance to
the Galactic center}. An exponential distribution is also adopted
for the vertical distribution of the thick disc, $\dfrac{dN}{dz}
\sim \exp{\dfrac{-z}{h}}$ with the scaleheight $h=900$ pc,
\textcolor{black}{$z$ is the distance to the Galactic plane}. For
the thin disc the logistic (sech-squared) distribution is
favoured: $\dfrac{dN}{dz} \sim \sech^2\dfrac{z}{2h}$. The thin
disc scaleheight $h$ of the mono-age population is expected to
increase with the stellar age \textcolor{black}{$t$}
\citep{2013ApJ...773...43B} as the stars form close to the
Galactic plane and then gradually migrate further from the
equator. The receipt from \citet{2003MNRAS.343.1231S} is adapted
\textcolor{black}{with the parameters $h_0 = 35$ pc, $h_1 = 78$ pc,
$v=175 \cdot 10^{-9} ~\rm{pc\cdot yr^{-1}}$, $\tau=10^9$ yr} (see
further discussion in Section \ref{modified_discuss}):

\begin{equation}
\label{eq:scaleheight}
h (\rm{t})=\begin{cases} \sqrt{(h_0^2+(tv)^2}~,~ t~  \le 0.5 \cdot 10^9 ~ \rm{yr}\\
 h_1 \sqrt{1+t/\tau}~,~ t~ > 0.5 \cdot 10^9 ~ \rm{yr}
\end{cases}
\end{equation}

The Sun is located in the Galactic plane at $r=R_{\odot}=8.5$ kpc from the Galactic center, its vertical offset is neglected. The azimuthal distribution at the galactic center $\dfrac{dN}{d\theta}$ is uniform. The distance to the stellar
system is calculated in cylindrical coordinates as
$d=r^2+R_{\odot}^2-2rR_{\odot}\cos{\theta}+z^2$. The maximum distance is set to $d_{max}=2$ kpc for practical reasons. Test simulations with $d_{max}=5$ kpc show that the fraction of binaries with $2~kpc<d<5$ kpc in the generated sample falls below
0.5 per cent for all reasonable parameter choices. $d_{max}=5$ kpc is used for the single-star population model. The brightest stars in the isochrones have absolute magnitudes around -10, allowing
them to be observed from larger distances; however, these stars are extremely rare and their contribution is small.

The interstellar extinction $A_v$ is estimated using the approximation formulas 12--13 from \citet{2016Ap.....59..548G}.
The model considers absorption in the Galactic equatorial and
Gould planes and it is expected to be reliable at least up to 2
kpc, thus safely covering the sample. $A_v$ is assumed to be equal
for both companions as the starlight passes through the same
interstellar medium. The apparent stellar magnitude is finally
calculated as $mag_{1,2}$=$Mag_{1,2}-5+5\log d+A_v$. $Mag_{1,2}$
is the absolute stellar magnitude derived from the isochrones. The
standard selective extinction factor $R_v=3.1$ is used to
calculate the $B_T$-band magnitude for single stars, interstellar
reddening caused by the excessive absorption in blue passband is
estimated as $E(B-V)=\dfrac{A_v}{R_v}=\dfrac{A_v}{3.1}$.

\subsection{Angular and projected separation}

Apart from $mag_{1,2}$, angular separation $\rho$ is a principal
observational parameter specific to visual binaries. The projected
separation between components is a product of $\rho$ and the solar
distance $d$: $s=\rho d$. \citet{1924PTarO..25f...1O} examined the
statistics of projected separations and showed that the variation
of frequency was low when using a logarithmic scale, implying
$\dfrac{dN}{ds} \sim s^{-1}$ with a proposed upper limit at 1 pc.
This conclusion remains plausible, though there are indications
that a steeper $dN/ds \sim s^{-\epsilon}, ~\epsilon \sim 1.5$
distribution is observed for $s > 2000 \sim 5000$ AU
\citep{2007AJ....133..889L,2012AJ....144..102T}.
\textcolor{black}{Notably $\epsilon =1.5$ is justified theoretically
as a consequence of wide binaries' formation during dispersion of
star clusters \citep{2020ApJS..246....4T}.} Some models prefer a
lognormal function \citep{2017MNRAS.464.4966H}. I adopt a broken
power law with the break point at $s = 5000 $ AU and
$\epsilon=1.6$ for large separations derived from
\citet{2017MNRAS.472..675A}:
\begin{equation}
    \label{eq:axis}
f(s) \sim
 \begin{cases}
0, s<s_{\rm min}, s>s_{\rm max} \\
s^{-\epsilon} & s\le5000 \text{ AU}\\
s^{-1.6} & s>5000 \text{ AU}\\
 \end{cases}
\end{equation}

$s_{\rm min}$ is set at 5 AU. According to the model, less than one per cent of the systems have $a<20~AU$; closer systems are
probably not evolutionary isolated and thus cannot be covered by this model. $s_{\rm max}=d_{max}\rho_{max}=30000$ AU \textcolor{black}{
if $d_{max}=2$ kpc is adopted. This value is well inside the Jacobi radius \citep{2010MNRAS.401..977J}, as only
massive and luminous pairs are observed from the large distances.
If originally formed, they are largely expected to survive in the
Galactic field.} Less than 5 per cent of the synthetic binaries
have $s>5000$ AU, the function is kept unaltered in this range,
while $\epsilon$ for closer separations is varied. The model
assumes that the projected distance distribution is independent
from the binary mass \textcolor{black}{and age. Orbital evolution
after the primary star leaves the main sequence can be
significant, however, the estimated mass loss does not exceed 2
per cent for 99 per cent of systems in the synthetic sample and does affect significantly very few of the sample objects.}

\section{Single-star model}
\label{single_sample}
\subsection{Standard model}
\label{standard_model}

The validity of the stellar population model is initially tested
on single stars. A comprehensive study of the single-star IMF is
beyond the scope of this paper, the single-star model is used
as a tool for verification of the Galaxy model. This
approach assumes that general properties, such as star formation
rate or spatial distribution, are identical for single and binary
populations. Their dynamical properties are different
\citep{1975MNRAS.173..729H}, e.g. the scaleheight potentially may
vary for single and binary stars, the model calibrated on single
stars is not necessarily the best choice for the binaries. Having
said this precaution, no corrections are applied,
single-star model parameters are used for the simulation of
binaries.

A magnitude-limited sample based on \textit{Tycho-2} catalogue is
used without additional modifications. \textit{Tycho-2} contains
120521 stars down to $mag=9$ in the $V_T$ passband and it is
expected to be nearly complete in this magnitude range
\citep{2000A&A...355L..27H}. Observational biases are anticipated,
e.g. due to significant presence of unresolved and resolved
binaries. Star counts $N$ predicted in the model
down to magnitude limits $mag=7 , 8, 9$ are calculated, then the
ratios $N_9/N_8$ , $N_8/N_7$ are compared to the corresponding
values in \textit{Tycho-2}. Additionally, the null hypothesis that
the distribution of $mag$ in observational and synthetic samples
arises from a common function is tested with the
Kolmogorov--Smirnov (KS) and Anderson--Darling (AD) two-sample
statistical tests. If the obtained $p$-value is less than 0.05
for either test, the result is considered negative. When $p$
exceeds this threshold for both the AD and KS tests, the null
hypothesis is not rejected and I consider the observational sample is
consistent with the model prediction. The size of the synthetic
sample is chosen empirically, the larger sample reduces the impact
of random fluctuations but takes more computational time. For the
single-star model, the synthetic sample is at least 3 times
larger than the observational one.

\begin{figure}
    \includegraphics[width=\columnwidth]{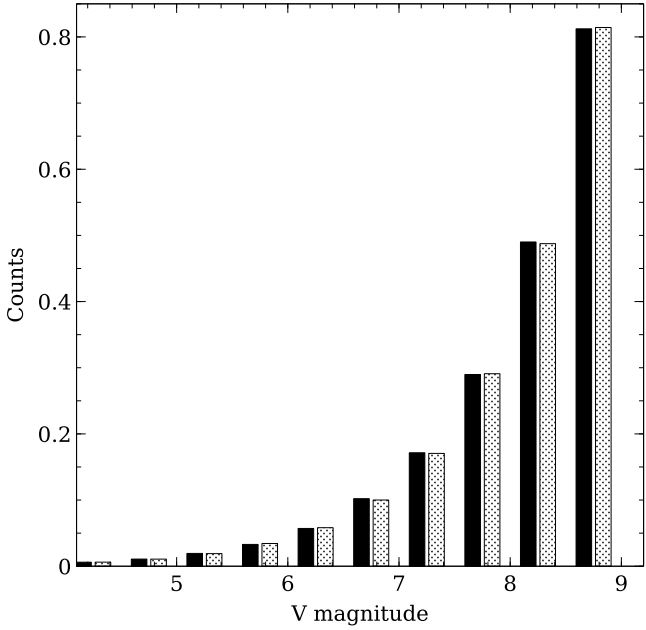}
    \caption{$V_T$ apparent magnitude distribution in \textit{Tycho-2} (dark) compared with the model prediction for single-star population.}
    \label{fig:visual}
\end{figure}

The predicted star counts depend on the chosen IMF slope. The best
agreement between the synthetic and observational samples is found
when $\alpha_2=2.6$, see Table \ref{tab:alpha} and Fig.
\ref{fig:visual}. A steep IMF overestimates the number of faint
stars; instead, smaller $\alpha_2$ causes an excessive abundance
of bright stars. Statistical tests agree with the basic $N_9/N_8$
comparison and favour $\alpha_2=2.6$. As the observational and
model samples are large ($N \sim 10^5$), the discriminatory power
of the tests is high, models with the adjacent values of
$\alpha_2= 2.5$ or 2.7 are statistically rejected. The IMFs with
$m_{\rm split}=0.5m_{\odot},~ m_{\odot}$ and universal
Salpeter-like $\alpha$ are considered; however, the difference is
barely noticeable as low-mass stars have a limited contribution to
the sample. The model produces the same number of stars with
$mag<9$ as in \textit{Tycho-2} when the mass density for
main-sequence stars, $m>m_{\odot}$, in the 50 pc solar radius is
$n=4.5 \cdot 10^{-3} \dfrac{m_{\odot}}{pc^3}$. This value is
10 per cent less than the mid-plane mass density
estimate by \citet{2017MNRAS.470.1360B}.

\begin{table}
    \centering
    \caption{Comparison of the \textit{Tycho-2} sample with the model predictions for single stars. The impact of varying $m_{\rm split}$ and $\alpha_2$ on the IMF (Eq. \ref{eq:imf}) is shown, other model parameters remain fixed. $\alpha_1=1.3$ for $m<m_{\rm split}$ when applicable. $N_{\rm mag}$ stands for star counts down to particular magnitude. The hypothesis that the magnitude distribution for the  observational and synthetic samples arises from a common distribution is not rejected by the AD and KS tests only for models with $\alpha_2=2.6$. * stands for \textit{Tycho-2} distances, corrected for unresolved binaries. See Section \ref{standard_model} for details.}
    \label{tab:alpha}
    \begin{tabular}{ccccccccc}
        \hline
         \multicolumn{2}{|c|}{IMF ($m^{-\alpha}$)}    & \multicolumn{2}{|c|}{Star counts}& \multicolumn{3}{|c|}{Distance quartiles, pc}\\
        \hline
        $\alpha_2$& $m_{\rm split}$& $N_9/N_8$  & $N_8/N_7$ & 25\% &50\% &  75\%\\
        \hline

        \multicolumn{2}{|l|}{\textit{Tycho-2}}&  2.87&2.95& 157 & 284 & 448\\
        \multicolumn{2}{|l|}{\textit{Tycho-2}*}& 2.87&2.95& 153 & 275 & 437\\
        \hline
        2.4 & $0.5m_{\odot}$ & 2.81 & 2.9 & 173 & 288 & 434\\
        2.5& $0.5m_{\odot}$& 2.84 & 2.91 & 169 & 284 & 426\\
        2.6& --& 2.87 & 2.95  & 165 & 280 & 418\\
        2.6& $0.5m_{\odot}$& 2.86 & 2.95 & 164 & 279 & 417\\
        2.6&$m_{\odot}$&2.87&2.98&166&280&416\\
        2.7&$0.5m_{\odot}$&2.90&3.00&161&277&412\\
        2.8&$0.5m_{\odot}$&2.92&3.04&157&273&406\\
        \hline
    \end{tabular}
\end{table}

\textit{Gaia} DR2 provides parallaxes for the majority of
\textit{Tycho-2} stars and allows to compare the distance
distribution in addition to the star counts. Since \textit{Gaia}
is strongly biased against bright stars, the cutoff at the bright
end is applied, leaving 106256 stars with $7<mag<9$ in the sample.
The best-neighbourhood search in the 3 arcsec radius with the
additional $G<11$ mag condition to exclude wrong matches finds
104559 \text{Gaia} DR2 entries with positive parallaxes. For the
795 stars with missing \textit{Gaia} data, \textit{Hipparcos}
parallaxes are available \citep{2007A&A...474..653V}, thus
providing distance estimate for more than 99 per cent of the
sample objects. Values of 25, 50, and 75 per cent distance
quartiles are compared for the model and the sample. The distance
to the sample stars is estimated as an inverse parallax $d=1/\pi$;
the use of \citet{2018AJ....156...58B} distances provides
marginally smaller values, e.g. shifting the median distance from
284 to 282 pc. A bias caused by the presence of unresolved
binaries in the catalogue is expected. In a magnitude-limited
sample, such systems are observed from a larger average distance
than normal single stars. A rough estimate considering  a 30 per
cent fraction of unresolved binaries and realistic $f(q)$ reduces
the median distance by around 10 pc. The effect is probably more
significant for distant stars, as massive bright stars are
generally expected to have larger multiplicity frequency.

The predicted and observed distance quartiles show moderate
agreement, see Table \ref{tab:alpha}. In addition to the general
sample, I separately consider samples near the equatorial plane
and the Galactic poles, see Table \ref{tab:alternatives}. The
model slightly underestimates the concentration of stars and
distances to far-away objects near the Galactic plane. This
discrepancy can be explained both by unaccounted biases in the
observational data and by model limitations. The synthetic sample
suggests that 18 per cent of stars in the solar neighborhood
belong to the thick disc population, in agreement with the local
normalisation estimate in \citet{2011MNRAS.414.2893F}.
Contribution of the thick disc stars to the synthetic sample is still low at around 7 per cent, thus undermining their impact.

\begin{table*}
    \centering
    \caption{Comparison of the \textit{Tycho-2} sample to the standard and modified models for single stars. All modified models use the same parameter values as in the standard model with the exception of the specified ones and $\alpha_2$ of the IMF (Eq. \ref{eq:imf}), which is adjusted for correct reproduction of \textit{Tycho-2} star counts. Distance quartiles are shown separately for the equatorial plane and the Galactic poles area, distances are inferred as inverse parallaxes. Closer stars are less prone to parallax-related biases and therefore larger inconsistency for 75 per cent quartile in comparison to 25 is allowed. F(\%) refers to the fraction of stars in the equatorial and polar zone, $\overline{B-V}$ is the median of the $B_T-V_T$ color distribution. See the discussion of the particular models in Section \ref{modified_discuss}.}
    \label{tab:alternatives}
    \begin{tabular}{ccccccccccccc}
        \hline
        Model & IMF&$N_9/N_8$ & \multicolumn{5}{c}{Equator, $|b|<5 \deg$}&\multicolumn{5}{c}{Pole, $|b|>60 \deg$}\\
         & $\alpha_2$&&F(\%)&\multicolumn{3}{c}{Distance,pc}&$\overline{B-V}$&F(\%)&\multicolumn{3}{c}{Distance, pc}&$\overline{B-V}$\\

        \hline
        Tycho 2 &&2.87&16.4&231&403&737&0.44&7.5&112&204&350&0.95\\

        \hline
    standard &2.6&2.86&15.4&225&370&631&0.35&8.3&116&216&336&0.98\\
    const SFR&2.9&2.86&15.3&214&348&580&0.30&7.9&113&201&319&0.69\\
    $\gamma=0.15$&2.4&2.88&15.7&236&388&666&0.38&8.4&120&230&349&1.03\\
    $t_{\rm max}=12$ Gyr&2.6&2.87&15.6&226&370&627&0.34&8.1&116&215&332&0.95\\
     $[Fe/H]$=0&2.6&2.84&15.9&226&368&636&0.30&8.1&113&208&319&1.03\\
    no thick disc &2.5&2.85&15.8&232&384&654&0.32&8.1&119&220&338&0.99\\
    exponential&2.5&2.87&17.2&233&383&660&0.26&8.5&138&254&376&1.00\\
    h - 30\%&2.9&2.85&15.8&195&315&501&0.46&7.7&90&174&300&0.92\\
    h + 30\%&2.4&2.87&15.3&253&427&778&0.28&8.5&141&247&362&0.99\\
 exp, $h=200$ pc&2.4&2.87&13.3&196&324&567&0.71&8.1&111&224&349&0.69\\
     logistic, $h=120$ pc&2.3&2.86&12.1&215&359&684&0.69&7.9&112&207&316&0.65\\
    $A_v$ - 30\% & 2.4&2.86&19.0 & 278&468&823&0.22&7.2&121&224&342&0.96\\
    $A_v$ + 30\% &2.6&2.85&13.8& 205&329&547&0.40&8.9&116&218&333&1.00\\

       \hline
    \end{tabular}
\end{table*}

\begin{figure}
    \includegraphics[width=\columnwidth]{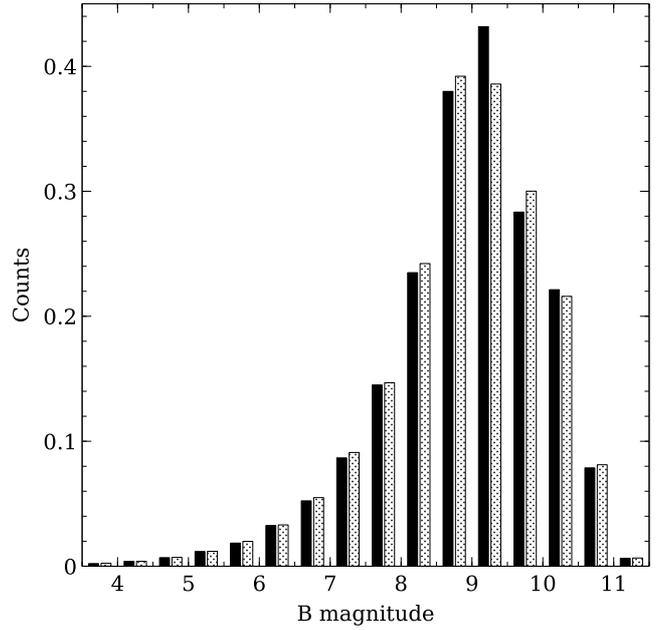}
    \caption{The $B_T$ apparent magnitude distribution
    for the restricted $V_T<9$ mag single-star sample (shown in dark color) compared to the model prediction.}
    \label{fig:blue}
\end{figure}

\begin{figure}
    \includegraphics[width=\columnwidth]{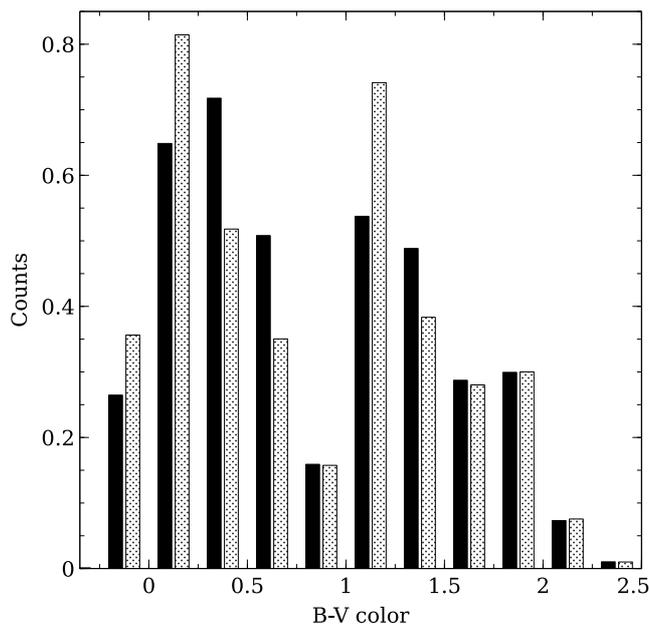}
    \caption{Comparison of the $B_T-V_T$ color distribution for single-star $V_T<9$ sample (dark color) to the model prediction.}
    \label{fig:color}
\end{figure}

The binary-star model focuses on $V_T$ photometry due to lacking multicolor photometry for binaries. The situation is better for single stars: for all but 82 entries, photometry in the \textit{Tycho}
$B_T$ passband is available. Distribution of the predicted and observed $B_T$ apparent magnitudes and $B_T-V_T$ colors is shown in Fig. \ref{fig:blue} and \ref{fig:color}. It shows a rather poor
agreement, and statistical tests actually consider them coming from different probability density functions. The \textit{Tycho-2}
color distribution is thoroughly studied and modelled in \citet{2014A&A...564A.102C}. I confirm that a very steep IMF with $\alpha_2 \sim 3 $ provides a better fit for the $B-V$ distribution; however, a shorter distance scale should be used to
keep star counts in agreement that drastically contradicts the \textit{Gaia} distances. Therefore, I adopt a model with more conventional $\alpha_2=2.6$, though it poorly fits the color
distribution. This value matches well with the \textit{Gaia}-based study by \citet{2019MNRAS.489.2377S}. Note that the IMF measured in the solar neighborhood is subject to observational biases increasing the observed $\alpha$ \citep{2018MNRAS.480.2449P}.

\subsection{Variations of the standard models}
\label{modified_discuss}

This section reviews the impact of particular parameters in the
standard model presented in Table \ref{tab:alternatives}. When a
specific parameter is changed, $\alpha_2$ is adjusted to ensure
that the star counts and $mag$ distribution are consistent with
the observational sample according to the statistical tests. IMF (Eq. \ref{eq:imf}) and SFR (Eq. \ref{eq:sfr}) are strongly correlated, as the steep IMF (large $\alpha$) increases the fraction of low-mass stars in the sample, similarly to the effect of high SFR in the past. Constant SFR ($\gamma=0$) induces a steep
IMF slope and a shorter distance scale in comparison to standard
model parameters ($\gamma=0.1~,~t_{\rm max}=13$ Gyr). Decreasing
SFR (larger $\gamma$) leads to a gentle IMF slope and larger
distances as old low-mass stars dominate the sample. Moderate
change of $t_{\rm max}$ essentially does not affect the results.
The impact of the metallicity is also limited, using the universal
solar metallicity does not significantly change the outcome.

Next, I consider the impact of spatial distribution (Section
\ref{distance}). No thick disc refers to a model adopting the thin
disc scaleheight law (Eq. \ref{eq:scaleheight}) for all ages,
while the standard model uses the distinct thick disc for
$t>10~Gyr$. The difference is subtle, as the thick disc
contribution is generally low. The thin disc dominates the sample
and makes vertical distribution law an important part of the
model. Exponential and logistic distributions were considered with
various age -- scaleheight dependences. In fact, none precisely
reproduces the distance distribution both for the Galactic poles
and equator, values in Eq. \ref{eq:scaleheight} are opted as the
best available approximation. The original formula from \citet{2003MNRAS.343.1231S}
\textcolor{black}{with the parameters $h_0 = 45$ pc, $h_1 = 177$ pc,
$v=417 ~\rm{pc\cdot yr^{-1}}$, $\tau=10^9$ yr} makes use of
exponential distribution, however, it poorly fits the distance
distribution near the Galactic poles, therefore it was adapted to
the logistic distribution. A scaled decrease of $h$ leads to a
steep IMF and a short distance scale, a larger value of scaleheight
works in the opposite direction. Notably, the distance
distribution alone can be adequately reproduced with a fixed
scaleheight independent from age, but such models predict a wrong
fraction of stars in the equatorial and polar zones and a wrong
color distribution. The interstellar extinction law is another
important factor, its adjustment significantly alters the
visibility of stars near the Galactic equator.

\section{Binary stars results}
\label{results}

Here I compare the observational sample obtained in Section
\ref{binary_sample} to the model predictions discussed in Section
\ref{model}. The routine is similar to the single-star model
validation in Section \ref{single_sample}, the important
difference is that binaries have more observational parameters,
while the observational sample is almost a hundred times smaller
than that for single stars. The general sample constrained by the
selection function (Eq. \ref{eq:selection})  contains $N_9=1227$
binaries; for $N_8=460$ of them, $mag_{1,2}<8$ and just for
$N_7=168$, both components are brighter than $mag=7$. Considering
that variance of the sample is a function of $\sqrt{N}$, the
random error around 3 per cent is expected for the general sample
and up to 8 per cent for the binaries with $mag_{1,2}<8$. A scarce sample size inevitably leads to a larger uncertainty of parameter estimates.

To minimize the effect of random fluctuations, the size of the
synthetic sample is chosen at least 20 times larger than that of
the observational data. The distributions of $mag_1$, $mag_2$,
$\Delta mag$, and $\rho$ are compared to the corresponding model
predictions using the KS and AD statistical tests; thereby, 8
separate tests are performed for each simulation, the $p$-value
0.1 is chosen as a threshold. Distance estimates are available for
all sample objects, see Section \ref{parallaxes}. Since individual
parallax values are prone to significant errors, distance
quartiles are compared. Star counts (number of binaries down to a
particular magnitude) and the distance distribution depend mainly
on the chosen IMF slope $\alpha_2$, see Table \ref{tab:RPT} and
\ref{tab:PCP+SCP}. The value of $\alpha_2$ is chosen to provide
consistency between the predicted and observed star counts. A
steep IMF (larger $\alpha_2$) overproduces faint stars and induces
a shorter distance scale, and vice versa.

\begin{table*}
    \centering
    \caption{Outcome of random pairing + twins pairing mechanism (RPT). The $f_{\rm twin}=0$ case is standard random pairing. The $p$-value refers to the lowest one among 8 separate results of the AD and KS statistical tests for $mag_1$, $mag_2$, $\Delta mag$, and $\rho$. While this value is not reproducible due to random fluctuations during generation, it clearly represents the agreement of the model with data, the larger $p$ means better fitting. The predicted star counts meet the  observational data when $\alpha_2=2$ is used. The reference local density $n$ for single stars is $45\cdot10^{-4} m_{\odot}/pc^3$. Usually distances are available for both components of the binary, quartiles for larger and smaller estimates are shown, see Section \ref{parallaxes}.}
    \label{tab:RPT}
    \begin{tabular}{cccccccccc}
        \hline
        scenario&IMF&\multicolumn{5}{c}{Distance quartiles, pc}&p-value&$N_9/N_8$&$n(M>m_{\odot})$\\
        $f_{\rm twin}$&$\alpha_2$& $\overline{dist_7}$ & $\overline{dist_8}$ & 25\% & 50\% & 75\%&&&$10^{-4}m_{\odot}/pc^3$\\
    \hline
   \multicolumn{2}{c}{Sample (min)} &89&106&77&132&231&&2.67 \\
   \multicolumn{2}{c}{Sample (max)} &97&109&81&139&251&&& \\
       \hline
       0&2.0&79&105&75&138&228&0&2.7&54\\
       0.025&2.0&85&114&81&149&248&0.06&2.68&47\\
       0.04&1.8&98&129&95&169&276&0.21&2.58&35\\
       0.04&2.0&90&116&82&151&252&0.49&2.66&44\\
       0.04&2.2&77&101&71&132&225&0.15&2.79&56\\
       0.055&2.0&94&121&85&157&261&0.11&2.64&41\\
       \hline
    \end{tabular}
\end{table*}

\begin{table*}
    \centering
    \caption{Outcomes of primary-constrained and split-core pairing (PCP and SCP) scenarios. A large set of values produces statistically acceptable results, a high $p$-value does not necessarily mean the best parameter choice. Note that the local mass density $n$ is largely dependant on the chosen twin excess $k_{\rm twin}$, while the distance distribution remains unaltered. See Fig. \ref{fig:results} and Section \ref{results} for the discussion. }
    \label{tab:PCP+SCP}
    \begin{tabular}{ccccccccccccc}
        \hline

scenario&$\beta$&$k_{\rm twin}$&$f_{\rm twin}$&$\alpha_2$& $\overline{dist_7}$ & $\overline{dist_8}$ & 25\% & 50\% & 75\% &p-value&$N_9/N_8$&$n(M>m_{\odot})$\\
       \hline
\multicolumn{5}{c}{Sample (min)} &89&106&77&132&231&&2.67&$10^{-4}m_{\odot}/pc^3$   \\
\multicolumn{5}{c}{Sample (max)} &97&109&81&139&251&&& \\
       \hline
PCP & 0 & 1&0 & 2.8 & 89&118&82&152&250&0&2.7&15 \\
PCP & -0.5 & 1&0 & 2.8 & 86 &115& 82&150&247&0.04&2.71&19\\
PCP & -1 & 1.1 &0.01& 2.6 & 93&122&89&160&261&0.05&2.62&22\\
PCP & -1 & 1.1 &0.01& 2.7 & 89 & 118& 85 & 155 & 255 &0.07&2.67&23\\
PCP & -1 & 1.1 &0.01& 2.8 &86&115&82&150&246&0.18&2.67&25\\
PCP & -1 & 1.1 &0.01& 2.9 &85&112&80&146&242&0.21&2.72&25\\
PCP & -1 & 1.1 &0.01& 3.0 &79&106&75&140&232&0.21&2.74&27\\
PCP & -1.5 & 1.4&0.07& 2.8&86&115&82&150&246&0.62&2.68&30\\
PCP & -2 & 1.8 &0.17& 2.6&90&120&89&159&260&0.15&2.64&41\\
PCP & -2 & 1.8 &0.17& 2.7& 89 &119& 86&154&251&0.15&2.64&44\\
PCP & -2 & 1.8 &0.17& 2.8& 84 &115& 84&151&247&0.15&2.66&44\\
PCP & -2 & 1.8 &0.17& 2.9&84&111&80&146&241&0.19&2.66&45\\
PCP & -2 & 1.8 &0.17& 3.0&80&109&78&142&236&0.06&2.73&48\\
SCP & 0 & 1 &0& 2.8 & 88&116&84&151&245&0.02&2.68&21\\
SCP & 0 & 1.1 &0.01& 2.8 & 88 & 117 & 83&153&249&0.13&2.69&20\\
SCP & -0.6 & 1.4 & 0.04 & 2.6 & 90 &122 &89&160&259&0.46&2.62&24\\
SCP & -0.6 & 1.4 &0.04& 2.7 & 87 & 117 & 84 &154 & 251 &0.51& 2.65 & 25\\
SCP & -0.6 & 1.4 &0.04& 2.8 & 85&115&82&151&248&0.58&2.69&28\\
SCP & -0.6 & 1.4 &0.04& 2.9 & 83&113&80&146&241&0.5&2.72&29\\
SCP & -0.6 & 1.4 & 0.04&3.0 &81&107&76&140&232&0.43&2.73&29\\
SCP & -1.4 & 2 &0.16& 2.8 & 90 &117&83&151&251&0.12&2.65&41\\
\hline
    \end{tabular}
\end{table*}

\begin{figure}
    \includegraphics[width=\columnwidth]{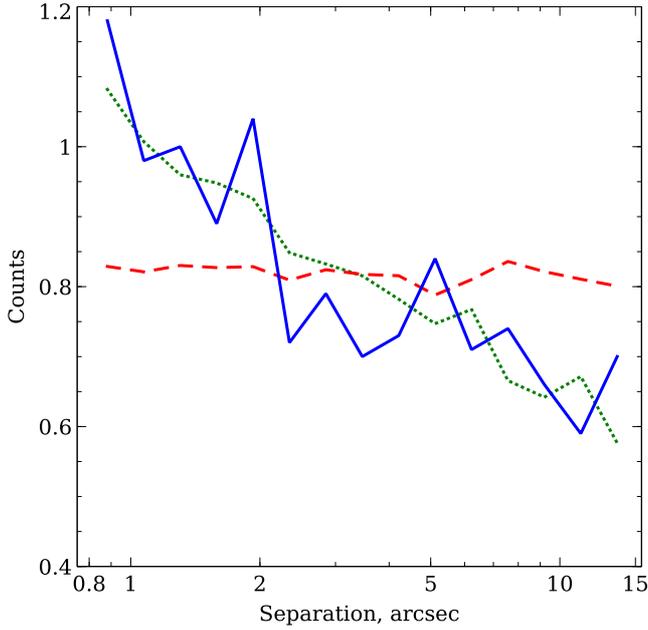}
    \caption{The distribution of angular separation ($\rho$) for the binary-star sample, the projected distance distribution slope $\epsilon$ in Eq. \ref{eq:axis} is varied. Solid blue: observational data; red dashed: $\epsilon=1$, green dotted: $\epsilon=1.2$. The latter value is clearly favoured by the statistical tests.}
    \label{fig:sep}
\end{figure}

It is noticed that the $\rho$ distribution is essentially
determined by the projected distance $s$ distribution and does not
depend on pairing scenarios or IMF choice. The synthetic and
observed $\rho$ distributions show the best agreement when
$\epsilon=1.2$ is adopted in Eq. \ref{eq:axis}, see Fig.
\ref{fig:sep}. The {\"O}pick law ($\epsilon=1$) clearly
underestimates the number of close binaries, statistical tests are
sharp and reject $\epsilon=1.1$ or 1.3. The angular separation is
a function of the distance from the Sun, $\rho=s/d$. While a
reasonable change of the model spatial distribution (Section
\ref{distance}) alters the preferred IMF slope, $\alpha_2$, $s$
remains largely unaffected. It is tempting to relate the deviation
from $\epsilon=1$ with the effect of dynamical evolution and
disruption of wide binaries. It should be kept in mind that 80 per
cent of the model binaries have $90<s<2000$ AU, therefore the
conclusion on preferred $\epsilon$ is valid for the limited range
of separations and stellar masses, see Table \ref{tab:summary}.

\begin{table}
    \centering
    \caption{Summary of the synthetic binary-star sample population. The average of the statistically acceptable models is shown.}
    \label{tab:summary}
    \begin{tabular}{cccc}
        \hline
        Parameter & 10\% & 50\% & 90\% \\
        \hline
        Primary mass $m_1, m_{\odot}$&1.1&2.2&4.4\\
        Secondary mass $m_2, m_{\odot}$&1&1.8&3.5\\
        Primary absolute magnitude&3.8&1&-1.1\\
        Secondary absolute magnitude&4.6&1.9&-0.2\\
        Mass ratio $q$& 0.65&0.89&0.99\\
        Age log $t$ (Gyr)&7.6&8.7&9.6\\
        Metallicity, $[Fe/H]$&-0.2&0&0.1\\
        Distance $d$, pc&45&151&372\\
        Extinction $A_v$, mag&0.07&0.20&0.52\\
        Projected separation $s$, AU&90&415&1940\\
        \hline
    \end{tabular}
\end{table}

\begin{figure}
    \includegraphics[width=\columnwidth]{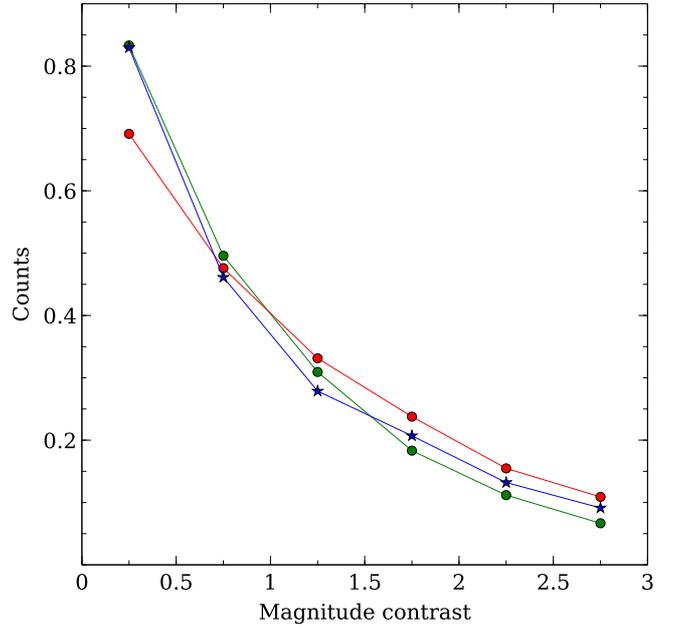}
    \caption{The magnitude contrast $\Delta mag =mag_2-mag_1$ distribution. Blue: observational sample; red: random pairing, $\alpha_2=2$; green: primary-constrained pairing, $\alpha_2=2.8$, $\beta=-0.5$, no twin excess. RP is clearly rejected due to a significant lack of binaries with equally bright companions in comparison with the observational data. PCP without twin excess fits the data better, however, the $p$-value $\sim 0.04$ is still below the threshold.}
    \label{fig:delta}
\end{figure}

Four pairing functions are examined in the paper: first, I
consider random pairing (RP) and random pairing + twins scenarios
(RPT), see Table \ref{tab:RPT}. The star counts meet the
observational data when $\alpha_2=2$ is adopted. However, the
$\Delta mag$ distribution shows a big disagreement as the model
significantly underproduces binaries with small magnitude
contrast, therefore random pairing is confidently ruled out, see
Fig. \ref{fig:delta}. The addition of a twin star population in the RPT algorithm improves the fit, the $p$-value of the statistical tests exceeds 0.1, and the model becomes consistent
with data when the adopted fraction of twin population is $f_{\rm
twin} \sim 0.03-0.055$. The distance distribution is also in a
reasonable agreement, though the observational scale is slightly
shorter than the model prediction. An important concern is the
produced local density of binary population. The obtained
mid-plane density for main-sequence single stars ($m>m_{\odot}$)
is $4.5 \cdot 10^{-3} \dfrac{m_{\odot}}{pc^3}$ (Section
\ref{standard_model}). The actual local density for binaries is
poorly constrained, however, it certainly does not exceed the
value for single stars. The inferred value for RPT is close to the
single-star figure, implying that nearly all $m>m_{\odot}$ stars
are binaries with a companion $m_2>m_{\rm min}=0.3m_{\odot}$,
which seems unlikely.

\begin{figure}
    \includegraphics[width=\columnwidth]{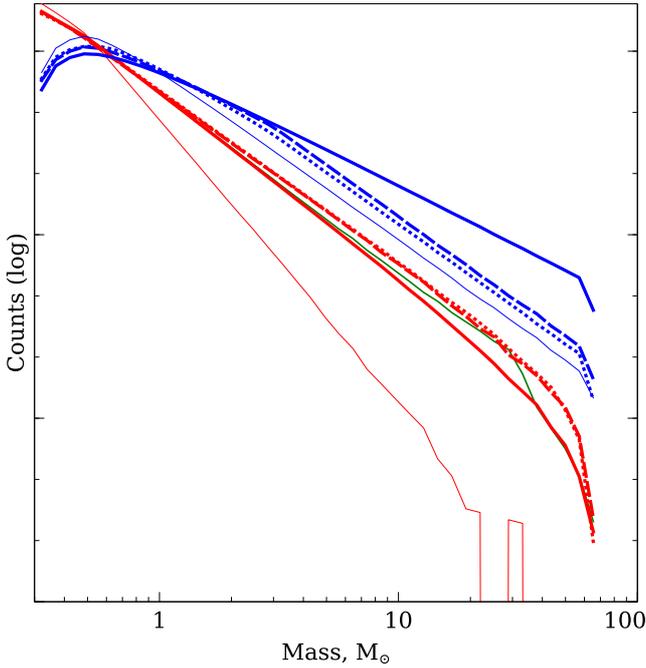}
    \caption{The produced IMF for primary (shown in blue) and secondary (red) components for several pairing mechanisms. Solid line: RP; thick: $\alpha_2=2$, thin: $\alpha_2=2.8$; dashed: PCP, $\beta=-1.5$, $k_{\rm twin}=1.5$; dotted: SCP, $\beta=-0.5$, $k_{\rm twin}=1.3$. Green: the secondary companion in RPT, $\alpha_2=2$, $f_{\rm twin}=0.04$, the primary overlaps the normal RP, shown in thick solid blue. The secondary companion mass distribution for RPT with $\alpha_2=2$ resembles PCP and SCP with $\alpha_2=2.8$ and determines the choice of the IMF slope.}
    \label{fig:m1+m2}
\end{figure}

Next, I proceed to primary-constrained pairing (PCP) and
split-core pairing (SCP). Both scenarios require IMF and f(q) (Eq.
\ref{eq:imf} and \ref{eq:q}) to be assigned. Models with a fixed
IMF but different $f(q)$ produce identical distance distributions
and star counts (Table \ref{tab:PCP+SCP}). Varying the IMF slope with a fixed $f(q)$ distribution, I favour $\alpha_2=2.8$ and perform further calculations for this value. Due to the small
sample size, the statistical tests tolerate at least $\Delta
\alpha_2=0.2$ error. The obtained value of IMF slope is notably different from the RPT scenario, which favours $\alpha_2=2$. The
secondary companion mass distribution appears to be critical for the observability of a binary system in the model (Fig. \ref{fig:m1+m2}). The $m_2$ distribution for RPT with $\alpha_2=2$ is similar to PCP and SCP with $\alpha_2=2.8$, matching the
difference of the preferred IMF slope.

\begin{figure}
    \includegraphics[width=\columnwidth]{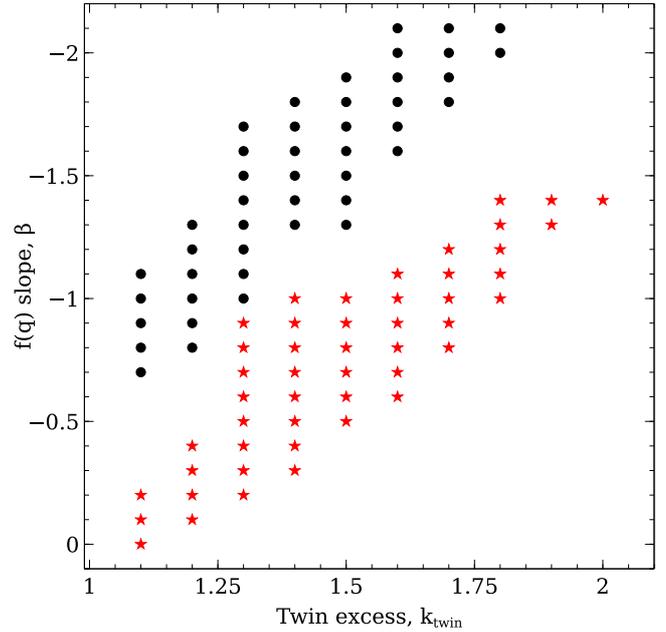}
    \caption{Values of mass-ratio distribution slope and twin excess in Eq. \ref{eq:q}, providing successful agreement ($p$-value > 0.1 in the AD and KS tests) of the observed and synthetic $mag_1$, $mag_2$ $\Delta mag$, $\rho$ distributions for the PCP (black) and SCP (red) pairing scenarios. A choice of a sole best-fitted model appears to be unjustified. Models with large $k_{\rm twin}$ require high binary frequency and local mass density, the latter parameter is poorly constrained. }
    \label{fig:results}
\end{figure}

Initially, models with a simple power slope without twin excess
are tested. The best agreement is found with f(q) slope $\beta=- 0.5-0.6 $ in
the case of PCP, but the obtained $p$-value for the $\Delta mag$
distribution is slightly below 0.05, and these models are statistically unlikely. Fig. \ref{fig:delta} shows that the
observed peak in the magnitude contrast distribution is sharper
than the predicted one. The distributions for $mag_1$, $mag_2$ and
$\rho$, at the same time, show a high $p$-value, emphasizing that
the rejection of the model is caused by poor fitting of $\Delta
mag$. Introduction of a small twin excess in $f(q)$ law improves
the fit, model prediction and observational data no longer diverge
statistically. A choice of a sole best-fitting combination of the
mass-ratio slope and twin excess is hardly justified, as the
statistical tests favour a reasonably wide set of parameters.
Pairs of $\beta$ and $k_{\rm twin}$ providing an acceptable
$p$-value larger than 0.1 are shown in Fig. \ref{fig:results} and
form two parallel bands for PCP and SCP. Models with a modest
mass-ratio distribution slope $\beta$ and small twin excess $k_{\rm twin}$ are
acceptable along with those having rapidly decreasing $f(q)$ with
a large $k_{\rm twin}$. Results for PCP and SCP are strongly
correlated, as expected, reflecting the connection discussed in
Section \ref{IMF_f(q)} and Fig. \ref{fig:PCP_SCP_RP}. While
uniform $f(q)$ ($\beta=0$) is confidently rejected for the PCP
scenario, it remains plausible for SCP with the addition of a
small twin excess, though it does not stand out among other viable
options.

The model comprises a purely binary population and does not
involve assumptions on the binary frequency. In Section
\ref{standard_model}, the local stellar density $n$ for
$m>m_{\odot}$ stars is used to validate the single-star model. The
produced $n$ for binaries potentially allows to discriminate
statistically consistent models. Indeed, $n$ depends on the $f(q)$
slope: while PCP models with $ \beta=-0.5$ imply $n \sim 2 \cdot
10^{-3} \dfrac{m_{\odot}}{pc^3}$, or around 45 per cent of the
single-star density, models with $\beta=-2$ require ubiquitous
binary frequency, see Table \ref{tab:PCP+SCP}. The
reported $n$ refers to the local density of primary stars;
inclusion of secondary stars with $m_2>m_{\odot}$ increases $n$
by approximately 15 per cent.  $n$ is estimated in assumption of
universal $\beta$ for the $0.1<q<0.95$ range; at the same time,
the model is not sensitive to systems with low and medium mass
ratios $q<0.6$, as they contribute less than 5 per cent of the
objects. If $f(q)$ flattens in the low-$q$ range, which is
plausible according to \citet{2017ApJS..230...15M}, $n$ decreases
and nearly 100 per cent binary frequency is no longer required. It
is widely accepted that the binary frequency increases with
stellar mass from roughly 0.5 for $m_1=m_{\odot}$ to 1 for the
most massive stars, thus favouring large $n$. At the same time,
the model does not cover the whole binary population, excluding
close binaries with $s<5$ AU, wide \textcolor{black}{$s>30000$} AU pairs, and
systems with low-mass companions $m_2<0.3 m_{\odot}$. The
mentioned factors obstruct the choice of the reference $n$;
however it is evident that scenarios requiring $n>4 \cdot 10^{-3}
m_{\odot}$ are not plausible.

The binary-star sample is constrained by the selection function,
Eq. \ref{eq:selection}. While it is hard to choose the sole
combination of parameters among wide range of options,
statistically acceptable models produce stellar populations with
similar properties, shown in Table \ref{tab:summary}. For example,
the median primary mass of the population is $2.2 m_{\odot}$, just
for 10 per cent of the objects $m_1<1.1 m_{\odot}$ or $m_1>4.4
m_{\odot}$. Efficiently, the conclusions on pairing mechanisms and
underlying distributions are valid for a limited range of
parameters. I find particularly instructive that, while the
synthetic sample with median $q= 0.89$ is dominated by the stars
with large $q$, the high fraction of twins in the overall
population is not necessary to explain the observed distribution.

Finally, I compare the results with those available in the
literature. \citet{2017ApJS..230...15M} compiled various sets of
observations, including  \citet{2002A&A...382...92S},
\citet{2010ApJS..190....1R} and \citet{2014MNRAS.437.1216D}, to
provide estimates of $\beta$ and $f_{\rm twin}$ separately for
different ranges of masses and orbital periods. My calculations
are made in terms of projected distance $s$, the exact relation of
$s$ and semi-major axis $a$ depends on the distribution of orbital
elements, including eccentricity and inclination. For this
purpose, it is enough to assume that the projected semi-major axis
$a$ is 10--20 per cent larger than $s$
\citep{1960JO.....43...41C,1969JRASC..63..275H}. Applying \textcolor{black}{Kepler's}
third law of motion $P \sim \sqrt{\frac{a^3}{m}}$, I estimate that
the majority of sample-star periods fall into 5 < log $P$ (days) <
7 category. According to \citet{2017ApJS..230...15M}, $f(q)$ slope
is drastically steeper for massive binaries and larger periods,
the presence of a twin-star population is noticeable only for
close and lower-mass binaries, see Table \ref{tab:comparison}.
However, the considered samples are small and the assessment of
the respective observational biases is challenging. My model
adopts universal $f(q)$ regardless of the orbital period, huge
variance of $\beta$ for the adjacent parameter domains in the
referenced paper is alarming and makes comparison ill-conditioned.
Marginalising \citet{2017ApJS..230...15M} results, small $f_{\rm
twin}$ with $\beta \sim -1.5$ can be expected, though the
uncertainty of $f(q)$ slope is high. My model predicts $f(q)$ with
a modest slope $\beta \sim -1$ with a low $f_{\rm twin}$. A
steeper $f(q)$ slope is also possible, but requires a larger twin
fraction.

\begin{table}
    \centering
    \caption{Excerpt of table 13 from \citet{2017ApJS..230...15M} and table G1 from \citet{2019MNRAS.489.5822E} covering the modelled range of primary masses, orbital periods (log $P$, days), or separations $s$ (AU) compared to the outcome of the PCP pairing mechanism. Three hypotheses on the fraction of twin stars are shown.}
    \label{tab:comparison}
    \begin{tabular}{cccc}
        \hline
        $m_1$, $m_{\odot}$ & log $P$ or $s$ &$\beta$&$f_{\rm twin}$\\
                \hline

         & log $P ~=~ 5$ & $-0.5\pm 0.3$ & $0.10\pm 0.03$\\
         & $50<s<350$ &$ -1.2 \pm 0.7$&$0.1\pm 0.03$\\
        0.8 --- 1.2 & $350<s<600$ &$ -1.8 \pm 0.4 $& $0.05 \pm 0.02$\\
         & $600<s<1000$ &$ -1.4 \pm 0.3 $&$ 0.04 \pm 0.02$\\
         & $1000<s<2500$ & $-1.5 \pm 0.2$ & $0.01 \pm 0.01$\\
         & log $P ~=~ 7$ &$-1.1 \pm 0.3$&$ <0.03$\\
                \hline
 & $50<s<350$ & $-0.9 \pm 1.0$&$0.09 \pm 0.08$\\
1.2---2.5& $350<s<600$ &$ -1.9 \pm 0.5 $&$ 0.09 \pm 0.05$\\
&$600<s<1000$& $-1.2 \pm 0.3$ & $0.02 \pm 0.03$ \\
&$1000<s<2500$&  $-1.6 \pm 0.2$ & $0.00 \pm 0.02$\\
                \hline

        2---5 & log $P ~=~ 5$ & $-1.4\pm 0.3$ & $<0.03$\\
        2---5 & log $P ~=~ 7$ & $-2.0\pm 0.3$&$<0.03$\\

                \hline
\multicolumn{2}{l}{PCP, small twin fraction}&$-0.9\pm 0.2$&0.01\\
\multicolumn{2}{l}{PCP, medium twin fraction}&$-1.5\pm 0.2$&0.07\\
\multicolumn{2}{l}{PCP, large twin fraction}&$-2.0\pm 0.1$&0.17\\

    \end{tabular}
\end{table}

The results are also compared to the \citet{2019MNRAS.489.5822E} study, based on \textit{Gaia} DR2
data, which provides a detailed coverage of the $f(q)$ dependence on projected separation. For solar-mass primaries,
\citet{2019MNRAS.489.5822E} clearly expect a steeper $f(q)$ than
\citet{2017ApJS..230...15M}, see Table \ref{tab:comparison}. No
clear trend for $\beta$ as a function of $s$ is seen, error
margins are significant. Notably, $\beta=-1.5$ is within one-sigma
error for all considered data domains, this fact probably
justifies the use of universal $\beta$ in my model. A twin excess
is noticeable for less massive and, as expected, closer binaries.
\citet{2019MNRAS.489.5822E} sample misses $m_1>2.5_{\odot}$ stars,
\citet{2017ApJS..230...15M} show that the $f(q)$ slope is steeper
for massive primaries, while $f_{\rm twin}$ is lower. Propagating
this trend to the missing domain in \citet{2019MNRAS.489.5822E}
leads to efficient $\beta \sim -1.5$ with small to moderate
$f_{\rm twin}$.

\section{Conclusions}
\label{conclusions}

After reviewing observational biases and adopting the selection
function (Eq. \ref{eq:selection}), an all-sky sample of 1227
visual binaries is created based on the WDS catalogue and supplied
with parallaxes from \textit{Gaia} DR2 and \textit{Hipparcos} in
Section \ref{binary_sample}. A population synthesis model adopts
various hypotheses about pairing mechanisms and fundamental
parameter distributions to transform them into apparent magnitudes
and angular separations in Section \ref{model}. The produced
distributions are tested against actual data using the Kolmogorov
-- Smirnov and Anderson -- Darling statistical tests. In Section
\ref{single_sample}, the model is calibrated on single stars,
favouring the single-star IMF slope $\alpha_2=2.6$ (Eq.
\ref{eq:imf}), though this value depends on the adopted SFR (Eq.
\ref{eq:sfr}), see Table \ref{tab:alpha} and
\ref{tab:alternatives}.

The main simulation results for binary stars are presented in
Section \ref{results}. It is noticed that the projected distance
distribution is largely independent of the assumptions on IMF,
$\epsilon=1.2$ is favoured in Eq. \ref{eq:axis}. Then, four
pairing functions are considered. Random pairing is confidently
rejected (Fig. \ref{fig:delta}), as it significantly underproduces
binaries with low magnitude contrast. A mix of random pairing with
a distinct twin population is statistically consistent when
$\alpha_2=2\pm0.2$ and $f_{\rm twin}\sim 0.04\pm0.015$, but
implies the entire population to be binary, which is presumed
unlikely, see Table \ref{tab:RPT}.

Next, primary-constrained and split-core pairing (PCP and SCP) with the universal power law $f(q)$ are considered. They provide a better agreement, but the produced $\Delta mag$ distribution is
still statistically rejected. However, introduction of a small twin excess creates favourable models (Table \ref{tab:PCP+SCP}). The inferred IMF slope for PCP and SCP $\alpha_2=2.8 \pm 0.2$ does not depend on the $f(q)$ choice. Combinations of acceptable mass-ratio function slope $\beta$ and twin
excess $k_{\rm twin}$ (Eq. \ref{eq:q}) are shown in Figure \ref{fig:results} and represent two parallel bands. SCP, which assumes the total system's mass to be fundamental, requires $\beta$ larger by $\sim
0.7$ in comparison to PCP. PCP considers the primary mass as a
principal parameter and favours $\beta \sim -1$, if the twin
fraction $f_{\rm twin}$ is as small as 0.01. A larger twin excess
is possible and induces a steeper $f(q) \sim q^{-1.5}$
distribution with a higher binary frequency. The latter factor
makes scenarios with large $f_{\rm twin}$ improbable.

The model successfully reproduces the observed $mag_{1,2}$,
$\Delta mag$, and $\rho$ distributions according to statistical
tests, distance estimates inferred from parallaxes are also in a
moderate agreement. The limits of the model reliability are
efficiently constrained by the observational sample. Although a
large grid of parameters is used in the input, characteristics of
most stars passed to the final synthetic sample lie in a fairly
bound parameter space. General properties of the synthetic
population are nearly invariant for all statistically acceptable
models, see Table \ref{tab:summary}. 80 per cent of the primary masses are in the $\sim$ 1 --- 4.5 $m_{\odot}$ interval,
the projected separation lies in the $\sim 10^2 - 2\cdot10^3$ AU
range. Only 10 per cent of systems have $q<0.65$. These values
highlight limitations of the model, conclusions on the pairing
function, IMF, $f(q)$, and $f(s)$ are valid for a specified range
of stellar masses and separations, as sensitivity of the model is
low outside it.

The results are relatively in line with the recent \citet{2017ApJS..230...15M} and \citet{2019MNRAS.489.5822E}
papers, though the obtained $f(q)$ slope is more modest, see Table
\ref{tab:comparison}. Contrary to this study, the referred papers
allow and report $f(q)$ depending on the projected distance or
orbital period. My model adopts a universal $f(q)$ and
still successfully reproduces the observed distributions. This
fact does not mean that stellar masses and separations are
independent, but implies that such approximation works fairly well
for the concerned parameter space. Caution is needed, as the
authors use slightly different parametrization for the twin
excess. Unlike the studies referred to, the present sample is
obtained without use of a parallax cutoff or \textit{Gaia's}
photometric data, thus avoiding the risk of related systematic
biases.

Finally, it should be acknowledged that the described population
synthesis model involves a number of simplifying assumptions, both
at the stage of sample assessment and numerical simulations. Some
of them are disputable, if not wrong. Those include treatment of
multiple stars, contact and semidetached components, while the
evolutionary model considers isolated stars. Photometry quality
for the sample is questionable, a dedicated multicolor survey is
crucial for a proper further analysis. The chosen pairing
mechanisms are basic and do not depend on stellar mass, age, or
orbit size. They successfully reproduce the fairly small
observational sample, but it is probable that a larger multicolor
dataset will require a more sophisticated approach.

\section*{Acknowledgements}
This research has made use of the SIMBAD database
\citep{2000A&AS..143....9W}, TOPCAT software
\citep{2005ASPC..347...29T}, and NASA's Astrophysics Data System
Bibliographic Services. The author thanks Dana Kovaleva,
Konstantin Malanchev, Oleg Malkov, \textcolor{black}{Nikolay Samus}, Alexey Sytov, and INASAN staff
for their support. \textcolor{black}{The referee's comments
allowed me to improve the paper.} The work was supported by the
Russian Foundation for Basic Researches (project 19-07-01198).

\section*{Data availability}
The data and code underlying this article are available in the
GitHub Repository at
\hyperlink{https://github.com/chulkovd/synthesis}{https://github.com/chulkovd/synthesis}.

\bibliographystyle{mnras}
\bibliography{chulkov.bib}

\bsp    
\label{lastpage}
\end{document}